\documentclass[12pt,preprint]{aastex}
\usepackage{amsmath}
\usepackage{ulem}

\shorttitle{Jet properties of $\gamma$-ray AGNs}
\shortauthors{Liang Chen}

\slugcomment{ApJS in press}

\begin{document}
\title{On the jet properties of the $\gamma$-ray loud active galactic nuclei}

\author{Liang Chen{$^{1,2}$}}

\altaffiltext{1}{Key Laboratory for Research in Galaxies and
Cosmology, Shanghai Astronomical Observatory, Chinese Academy of
Sciences, 80 Nandan Road, Shanghai 200030, China; E-mail: chenliang@shao.ac.cn}
\altaffiltext{2}{University of Chinese Academy of Science, 19A Yuquanlu, Beijing 100049, China}


\begin{abstract}

Based on broadband spectral energy distribution (SEDs), we estimate the jet physical parameters of 1392 $\gamma$-ray-loud active galactic nuclei (AGNs), the largest sample so far. The (SED) jet power and magnetization parameter are derived for these AGNs. Out of these sources, the accretion disk luminosity of 232 sources and (extended) kinetic jet powers of 159 sources are compiled from archived papers. We find the following, (1) Flat-spectrum radio quasars (FSRQs) and BL Lacs are well separated by $\Gamma=-0.127\log L_{\gamma}+8.18$, in $\gamma$-ray luminosity versus photon index plane with a success rate of 88.6\%. (2) Most FSRQs present a (SED) jet power larger than the accretion power, which suggests that the relativistic jet-launching mechanism is dominated by the Blandford-Znajek process. This result confirms previous findings. (3) There is asignificant anticorrelation between jet magnetization and a ratio of the (SED) jet power to the (extended) kinetic jet power, which, for the first time, provides supporting evidence for the jet energy transportation theory: a high-magnetization jet may be more easily to transport energy to a large scale than a low-magnetization jet.


\end{abstract}


\section{Introduction}
\label{sec:introduction}
The gravitational potential of the suppermassive black hole (BH), at the center of an active galactic nucleus (AGN), is believed to be the ultimate energy source of the AGN. An accretion disk can be formed during way of matter falling onto the BH, and the angular momentum can be lost through viscosity or turbulence  \citep[e.g.,][]{1984ARA&A..22..471R} or via outflow or magnetic field processes \citep[e.g.,][]{2011ApJ...737...94C, 2013ApJ...765..149C}. About 10\% of AGNs have relative stronger radio emissions compared with their optical emissions (i.e., radio-loud AGNs\footnote{The radio-loudness parameter, the ratio of the radio flux at
5 GHz to the optical flux at the $B$ band, $R\equiv f_{\rm 5 GHz}/f_{B}>10$ for radio-loud AGNs \citep{1989AJ.....98.1195K}.}), which are believed to host relativistic jets launched from the central accreting system \citep{1995PASP..107..803U, 2008ApJ...685..801Y, 2016ApJ...833...30C}. The jet launching can be related to the Blandford-Znajek (BZ) process \citep[][]{1977MNRAS.179..433B} through extracting the rational energy of BHs and/or the Blandford-Payne (BP) process \citep[][]{1982MNRAS.199..883B} by releasing the gravitational energy of the accretion disk. As an extreme subclass of radio-loud AGNs, blazars show broadband emissions (radio through $\gamma$-ray), rapid variability, high and variable polarization, superluminal motions, and core-dominated nonthermal continua, which are believed to be caused by a relativistic Doppler-beaming effect due to a small viewing angle between the relativistic jet and the line of sight \citep[see, e.g.,][]{1978bllo.conf..328B, 1995PASP..107..803U, 2014A&ARv..22...73F, 2016ARA&A..54..725M}. These properties provide the blazar an ideal laboratory to study AGN jet physics. Broadband spectral energy distributions (SEDs) of blazars usually show two significant bumps: one peaks at the infrared (IR) to X-ray bands, which is believed to be the synchrotron emission of energetic electrons within the jet; and the second bump peaks at the $\gamma$-ray band, which may be the inverse Compton (IC) emission of the same electron population emitting the synchrotron bump \citep[e.g.,][see also the hadronic model, \citet{1993A&A...269...67M, 2000NewA....5..377A}]{1996ApJ...461..657B, 1994ApJ...421..153S, 1993ApJ...416..458D}. According to the rest-frame equivalent width (EW) of the broad emission lines, blazars are classified as flat-spectrum radio quasars (FSRQs) with EW$\geq5${\AA}) and BL Lacertae objects (BL Lacs) with EW$<5${\AA} \citep[see][]{1997A&A...325..109S}.

Indisputably, the jet property is basically important in studying jet physics, including the jet launching, energy transportation, energy transform and conversion, radiative processes, etc. Limited by spatial resolution of modern telescopes, only the very long baseline interferometry (VLBI) radio technique can resolve subparsec-scale jet \citep[even down to several Schwarzschild radii for very near and massive BHs, e.g., M87,][]{2012Sci...338..355D}. Superluminal motion is a common phenomenon in blazar VLBI observations, which set a strong constraint that these jets should move very fast \citep[apparent velocity on the order of $\beta_{\rm app}\equiv v_{\rm app}/c\gtrsim$10, and can reach $\sim$30 for some special sources; see, e.g.,][]{2013AJ....146..120L}. The rapid variability at the $\gamma$-ray band can also set a lower limit of jet velocity due to the fact that these $\gamma$-ray photons are actually escaped from the emission region by overcoming the possible absorption of soft photons through photon annihilation \citep[e.g.,][]{1995MNRAS.273..583D}. The jet properties at the launching and energy dispassion regions can also be constrained through broadband SED modeling \citep[e.g.,][]{1998MNRAS.301..451G, 2014Natur.515..376G, 2015MNRAS.448.1060G, 2017ApJ...842..129C}. This powerful method can offer a constraint on the main jet parameters, including, within the one-zone leptonic model, the Doppler factor (the bulk Lorentz factor sometimes), the strength of the magnetic field, the energy density of the relativistic electrons, the jet power, the energy distribution of the electrons, and the location of the emission region \citep[e.g.,][]{2014Natur.515..376G, 2015MNRAS.448.1060G, 2014ApJS..215....5K, 2017ApJ...842..129C}. Benefit from a better knowledge of the high-frequency radio, millimeter, far-IR, and $\gamma$-ray continuum given by the {\it WMAP}, {\it Planck}, {\it WISE} and {\it Fermi} Large Area Telescope (LAT) satellites, \citet{2014Natur.515..376G} modeled the broadband SED of the largest blazar sample yet \citep[the number of blazars reaches 217; see also][]{2015MNRAS.448.1060G} and got the jet physical parameters. Note that within the one-zone leptonic model, the estimation of the jet parameters mainly depends on the peak frequency and luminosity of the two SED bumps \citep[see, e.g.,][]{1998ApJ...509..608T, 2017ApJ...842..129C}.

Since its launch on 2008 June 11, the LAT onboard the {\it Fermi} satellite has revolutionized our knowledge of the $\gamma$-ray AGNs above 100 MeV. A large number of AGNs are detected by {\it Fermi}/LAT and are compiled as the LAT Bright AGN Sample \citep[LBAS;][]{2009ApJ...700..597A} and the First/Second/Third LAT AGN Catalogs \citep[1LAC/2LAC/3LAC;][]{2010ApJ...715..429A, 2011ApJ...743..171A, 2015ApJ...810...14A}. The 3LAC includes 1591 AGNs located at high Galactic latitudes ($\mid b\mid>10^{\circ}$) and detected at $\gtrsim100$ MeV with a test statistic greater than 25 between 2008 August 4 and 2012 July 31. The number of $\gamma$-ray AGNs is even beyond the previous estimation \citep[e.g.,][]{2008ApJ...673L.131C}. Most of these AGNs are blazars (98\%) and can be obtained with radio, optical, and X-ray data. In this paper, we will investigate the AGN jet physics based on broadband SED. The Section \ref{sec:sample} describes the sample. The Section \ref{sec:method} presents the method to calculate the jet parameters. The result and discussion will be given in the Section \ref{sec:resultdiscussion}. Our conclusion will be drawn in the Section \ref{sec:conclusion}. A $\Lambda$CDM cosmology with values from the Planck results is used in our calculation; in particular, $\Omega_{m}=0.32$, $\Omega_{\Lambda}=0.68$, and the Hubble constant $H_{0}=67$ km s$^{-1}$ Mpc$^{-1}$ \citep{2014A&A...571A..16P}.

\section{Sample}
\label{sec:sample}

The 3LAC contains 1591 AGNs, of which 1559 are blazars (467 FSRQs, 632 BL Lacs, and 460 BCUs\footnote{BCU refers to the blazar candidates of uncertain type \citep{2015ApJ...810...14A}.}). Due to the uniform all-sky exposure of the {\it Fermi}/LAT, these sources form a $\gamma$-ray flux-limited sample. Note that since the blazars are rapidly variable, it is better to use the simultaneous multifrequency data, which seems impossible for the 3LAC, given the very large number of sources. Through collecting the multifrequency data (radio to X-ray bands) from the NASA/IPAC Extragalactic Database (NED), \citet{2016ApJS..226...20F} successfully obtained the broadband SED of 1392 blazars (461 FSRQs, 620 BL Lacs, and 311 BCUs). They employed a log-parabolic function, $\log(\nu F_{\nu})=-P_{1}(\log\nu-\log\nu_{p})^{2}+\log(\nu_{p} F_{\nu_{p}})$, to fit the SED; therefore, the synchrotron peak frequency ($\nu_{p}$), spectral curvature ({$P_{1}$}), and the peak flux ($\nu_{p} F_{\nu_{p}}$) are obtained. Of the total number of blazars, 999 have measured redshifts. In this paper, we will study the jet physical properties of the 1392 blazars, based on the data of \citet{2016ApJS..226...20F}. The data regarding the $\gamma$-ray (i.e., photon index and luminosity) are compiled from \citet{2015ApJ...810...14A} and \citet{2015ApJS..218...23A}.

\section{Methods}
\label{sec:method}

\begin{figure}
\begin{center}
{\includegraphics[width=1.0\linewidth]{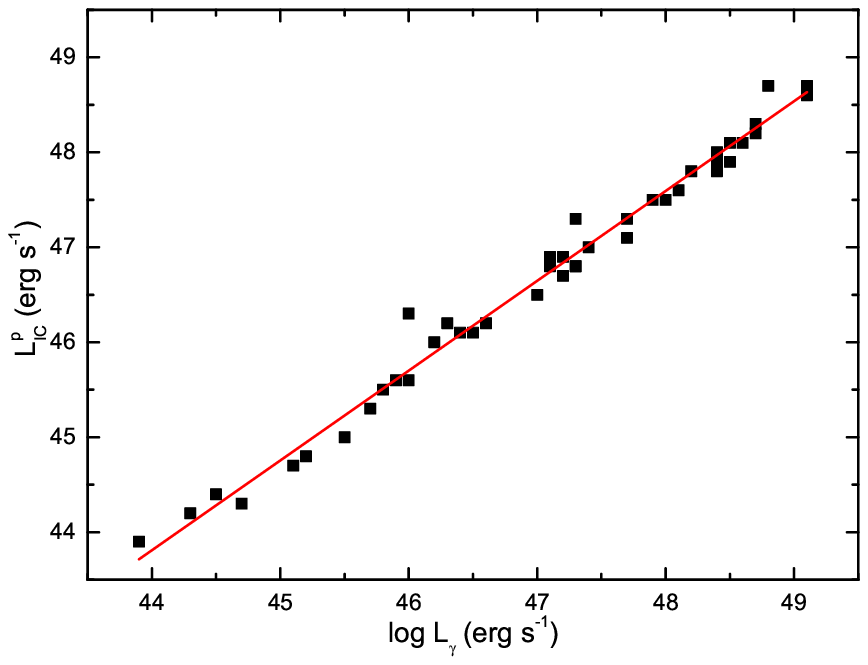}}
\end{center}
\caption{Relation between $\gamma$-ray luminosity and IC peak luminosity for blazars from \citet{2010ApJ...716...30A}, with the best linear fit $\log L_{IC}^{p}=(0.946\pm0.018)\log L_{\gamma}+(2.18\pm0.88)$ and the chance probability $p=8.72\times10^{-39}$ (Pearson test).}
\label{fig:luminosity_g_p}
\end{figure}

We directly collect the peak frequency and flux of the synchrotron bump from \citet{2016ApJS..226...20F}. For the IC bump, the observed $\gamma$-ray luminosity and photon index are used to estimate the peak luminosity and frequency. \citet{2010ApJ...716...30A} collected quasi-simultaneous broadband SEDs of 48 blazars and employed a third-degree polynomial to fit the synchrotron and IC bumps. The obtained peak frequency and luminosity are found to be significantly correlated with the $\gamma$-ray photon index and $\gamma$-ray luminosity, respectively. They found a tight relation between the photon index and IC peak frequency: $\log\nu_{IC}^{p}=-4\Gamma+31.6$ \citep[see Equation 5 in][]{2010ApJ...716...30A}. For the same sample, we plot $\gamma$-ray luminosity versus IC peak luminosity in Figure \ref{fig:luminosity_g_p}. The linear fitting shows $\log L_{IC}^{p}=(0.946\pm0.018)\log L_{\gamma}+(2.18\pm0.88)$ and the Pearson test yields the chance probability $p=8.72\times10^{-39}$. Because there is not enough data to construct the IC bump for this large sample, the above two formulae are used to estimate the IC peak frequency and luminosity.

\begin{figure}
\begin{center}
{\includegraphics[width=1.0\linewidth]{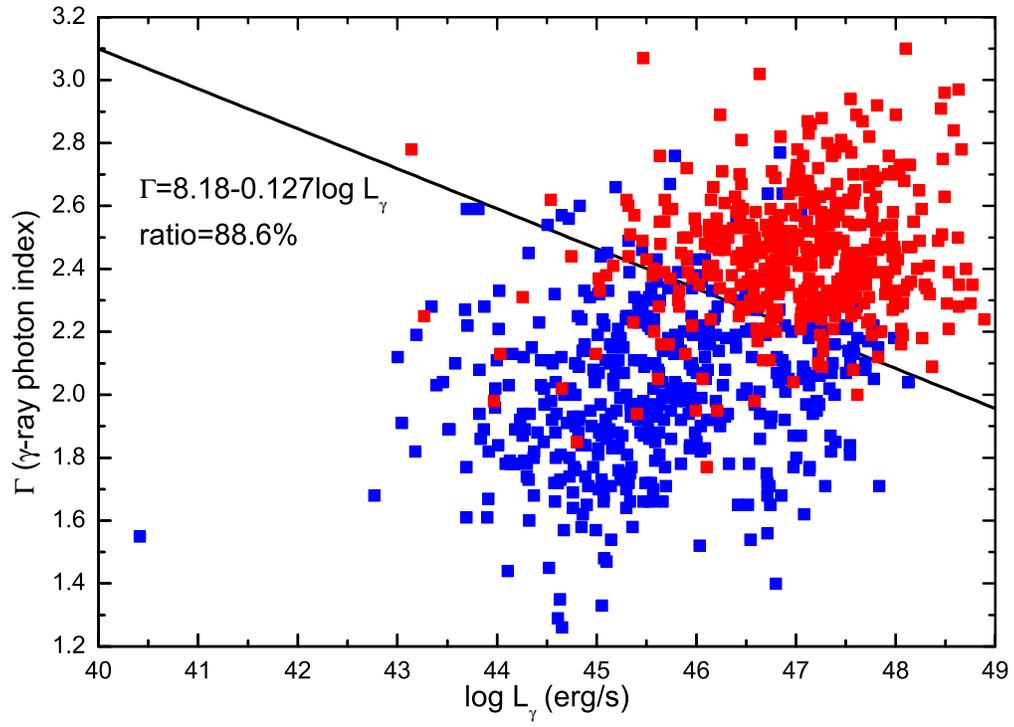}}
\end{center}
\caption{The $\gamma$-ray photon index versus $\gamma$-ray luminosity. The FSRQs and BL Lacs are well separated by $\Gamma=-0.127\log L_{\gamma}+8.18$, with a success rate of 88.6\%.}
\label{fig:blazar_divide}
\end{figure}

For BL Lac objects, the optical emission lines are very weak or even missed. Their $\gamma$-ray emission is believed to be synchrotron self-Compton (SSC) emission. The $\gamma$-ray emission of FSRQs, with strong optical emission lines, is believed to be emission from IC external seed photons produced (EC) from, e.g., a broad emission-line region (BLR) or dust torus. As suggested by \citet{2009MNRAS.396L.105G}, BL Lacs and FSRQs are neatly separated in the $\gamma$-ray photon index versus $\gamma$-ray luminosity plane \citep[see also][]{2009ApJ...700..597A, 2015ApJ...810...14A}. For confirmed FSRQs and BL Lacs in the 3LAC \citep{2015ApJ...810...14A}, we plot the $\gamma$-ray photon index versus $\gamma$-ray luminosity in the Figure \ref{fig:blazar_divide}. It can be seen that FSRQs have larger $\gamma$-ray luminosity and softer $\gamma$-ray spectra (red squares, mostly in the upper right corner) compared with BL Lacs (blue squares, mostly in the lower left corner). The physical difference between these two subclasses may be related to the different accretion model of the central BH; i.e., a standard cold accretion disk exists in FSRQs, while advection-dominated accretion flow \citep[ADAF;][]{2014ARA&A..52..529Y} exists in BL Lacs \citep[e.g.,][]{2002ApJ...579..554W, 2003ApJ...599..147C, 2009ApJ...694L.107X, 2009MNRAS.396L.105G, 2014MNRAS.445...81S}. Despite of the different physical origin, we choose a phenomenological criterion (a line in the $\gamma$-ray photon index versus luminosity plane) to roughly separate these two subclasses. This criterion/line should satisfy two criteria (see the Figure \ref{fig:blazar_divide}): (1) the fraction of FSRQs above the line and the fraction of BL Lacs below the line should be same, and (2) this fraction value must reach maximum. Finally, we get the criterion (the black line in the Figure \ref{fig:blazar_divide}): $\Gamma=-0.127\log L_{\gamma}+8.18$. In this case, the fraction of FSRQs above the line and the fraction of BL Lacs below the line reaches the maximum value: 88.6\%. As mentioned above, 311 out of 1932 blazars are BCUs. In this paper, this criterion is employed to classify these BCUs into FSRQs or BL Lacs. In this case, the class type is labeled ``UF" for BCUs classified as FSRQs and ``UB" for BCUs classified as BL Lacs, while, the class types ``CF" and ``CB" are for these confirmed FSRQs and BL Lacs, respectively (see the class type in Table \ref{tab:data}).

The SSC emission of FSRQs usually peaks at the X-ray band and dominates its X-ray emission \citep[see, e.g.,][]{2014ApJS..215....5K, 2015ApJ...807...51Z, 2015MNRAS.448.1060G}. Therefore, the X-ray luminosity can be set as the upper limit of the SSC luminosity in FSRQs. \citet{2016ApJS..226...20F} presented X-ray luminosity at 1 keV for 275 out of 486 FSRQs and found a tight relation between X-ray and $\gamma$-ray luminosity, $L_{\rm 1GeV}=0.91L_{\rm 1keV}+5.1$ \citep[$p<10^{-4}$; see Table 3 in][]{2016ApJS..226...20F}, which is used to estimate the X-ray luminosity for FSRQs without X-ray measurement.

In our sample, 393 sources have no measured redshift (157 BL Lacs and 236 BCUs). For these sources, we use the median values of the known redshifts of confirmed BL Lacs ($z=0.359$) and confirmed FSRQs ($z=1.108$) for unknown-redshift BL Lacs (including BCUs divided into BL Lacs) and BCUs divided into FSRQs, respectively. Finally, we have 803 BL Lacs and 589 FSRQs, of which the class type, redshift, and other parameters are presented in Table \ref{tab:data}.

We note that there is only a blazar having extremely large synchrotron peak frequency above $>10^{20}$ Hz in the sample: 3FGL J1504.5-8242 (1RXS J150537.1-824233) having $\log\nu_{\rm s}^{\rm p}(\rm Hz)=20.12$, which is a BCU \citep{2015ApJS..218...23A, 2016ApJS..226...20F}. It has a $\gamma$-ray photon index $\Gamma_{\gamma}=2.3$ \citep{2015ApJS..218...23A}, which is a typical value of FSRQs. According to the above criterion, this source is classified as a FSRQ. Its synchrotron peak frequency may be overestimated because of that the quality of its synchrotron broadband SED is bad\footnote{http://ned.ipac.caltech.edu/}. For this source, the synchrotron peak frequency is assumed to be an average value of FSRQs: $\log\nu_{\rm s}^{\rm p}(\rm Hz)=13.94$ \citep[see][]{2016ApJS..226...20F}.

A one-zone synchrotron + IC model is adopted in this paper to estimate the jet physical parameters. This model assumes a homogeneous and isotropic emission region, which is a sphere with radius $R$, a uniform magnetic field with strength $B$, and a uniform electron energy distribution $N(\gamma)$. The emission region moves relativistically with a Lorentz factor $\Gamma=1/\sqrt{1-\beta^{2}}$ and a viewing angle $\theta$, which forms the Doppler factor $\delta=1/\left[\Gamma\left(1-\beta\cos\theta\right)\right]$. The frequency and luminosity transform from jet to AGN frames as\footnote{The prime refers to values measured in the jet frame.}, $\nu=\delta\nu'$ and $\nu L(\nu)=\delta^{4}\nu' L'(\nu')$, respectively. As stressed in \citet{1998ApJ...509..608T}, the jet parameters specifying the one-zone model are uniquely determined once the basic observables and variability time-scale are known. Therefore, along the lines of the analytical treatment in \citet{1998ApJ...509..608T} for the one-zone SSC model, it is possible to derive a useful approximate analytical expression for the jet physical parameters as a function of the observed SED quantities (e.g., the peak frequency and luminosity of the synchrotron and IC components). Because of that the synchrotron emission at the peak frequency are usually optical thin, we have
\begin{equation}
L_{sy}'(\nu')=\frac{4}{3}\pi R^{3}\cdot4\pi j(\nu'),
\label{equ:lum}
\end{equation}
where the emitting efficient is
\begin{equation}
j_{sy}(\nu')=\frac{1}{4\pi}\int N(\gamma)P_{sy}(\nu',\gamma)d\gamma.
\label{equ:jem}
\end{equation}
Here $P_{sy}(\nu',\gamma)$ is the synchrotron emission power of a single electron. Because of that the SED of the synchrotron emission of a single electron is very narrow in the frequency space \citep[see, e.g.,][]{1970RvMP...42..237B, 1979rpa..book.....R}, one sometimes assumes that all emission is produced at a particular frequency, i.e., monochromatic approximation \citep[see][for detail]{2017ApJ...842..129C}, using the following equation as an approximation of $P_{sy}(\nu',\gamma)$,
\begin{equation}
P_{sy}(\nu',\gamma)\approx\frac{4}{3}\sigma_{T}cU_{B}\gamma^{2}\delta(\nu'-\frac{4}{3}\nu_{L}\gamma^{2}),
\label{equ:Psy}
\end{equation}
where $\nu_{\rm L}=eB/2\pi m_{\rm e}c$ is the Lamer frequency and $U_{\rm B}=B^{2}/8\pi$ is the magnetic field energy density. In this case, the synchrotron luminosity (Equation \ref{equ:lum}) will be reduced to
\begin{equation}
L_{sy}'(\nu')=\frac{4}{3}\pi R^{3}\cdot4\pi\cdot\frac{\sigma_{\rm T}c}{8\pi}\frac{U_{\rm B}}{\nu_{\rm L}}\gamma N(\gamma),
\label{equ:jem_delta}
\end{equation}
where $\nu'=(4/3)\nu_{L}\gamma^{2}$.

In leptonic blazar jet models, the synchrotron and IC emission components are radiations of nonthermal electron populations that are assumed to be isotropic in the jet fluid
frame. One technique is to fit the data by injecting power-law electron distribution and allowing the electrons to evolve in response to radiative and adiabatic energy losses \citep[e.g.,][]{2002ApJ...581..127B, 2003A&A...406..855M, 2003A&A...410..101K}. In this case, many parameters must be specified, including the cutoff energies, injection indexes, and power, but this method is potentially useful to follow the dynamic spectral behavior of blazars. Contrary to this approach, we abandon any preconceptions
about particle acceleration and employ the simplest functional
form that is able to provide reasonably good fits to the SED data \citep[see, e.g.,][for SSC modeling of TeV
blazars with power-law electron distributions]{2008ApJ...686..181F}. For this purpose, and with the goal of minimizing the number of free parameters, a three-parameter log-parabolic function is employed to describe electron energy distribution in this paper \citep[see][]{2014ApJ...782...82D, 2016ApJS..226...20F},
\begin{equation}
N(\gamma)=N_{0}\left(\frac{\gamma}{\gamma_{0}}\right)^{-3}10^{-b\log^{2}\left(\frac{\gamma}{\gamma_{0}}\right)}.
\label{equ:Ngamma}
\end{equation}
This electron energy distribution is only phenomenologically assumed to follow the log-parabola, without taking self-consistently into account the evolution due to injection and cooling effects \citep[e.g.,][]{1998A&A...333..452K, 2011ApJ...739...66T}. Despite this, the log-parabolic model can successfully represent the broadband SED of blazars in both observation \citep[single-source and sample studies; e.g.,][]{2004A&A...413..489M, 2014ApJ...788..179C, 2015ApJ...809..174D, 2016A&A...591A.130K, 2016MNRAS.463.3038X, 2016ApJS..226...20F} and theoretical SED modeling \citep[e.g.,][]{2011ApJ...739...66T, 2013ApJ...771L...4C, 2014ApJ...782...82D, 2015MNRAS.454.1310Y, 2017MNRAS.464..599D, 2017arXiv171105494H}. As for why there is a term $(\gamma/\gamma_{0})^{-3}$, it is to guarantee that the electrons of $\gamma_{0}$ emit at the peak frequency in the $\log\nu'-\log\nu' L'_{\nu'}$ frame (see below). The entire description of the electron energy distribution is then given by three parameters: the normalization $N_{0}$, the peak Lorentz factor $\gamma_{0}$ of nonthermal electrons, and the spectral curvature parameter $b$.

The curvature, $b$, of the electron energy distribution has a relation with that of the SED (synchrotron bump),
$b\approx5P_{1}$ \citep[see, e.g.,][]{2006A&A...448..861M, 2014ApJ...788..179C}. In this case, the synchrotron peak frequency in the $\log\nu'-\log\nu' L'_{\nu'}$ frame is
\begin{equation}
\nu'^{p}_{sy}\simeq(4/3)\nu_{L}\gamma_{0}^{2},
\label{equ:nu_sy}
\end{equation}
and the corresponding peak luminosity is
\begin{equation}
\nu'^{p}_{sy}L_{sy}'(\nu'^{p}_{sy})=4\pi\frac{16}{9}\pi R^{3}\frac{\sigma_{\rm T}c}{8\pi}U_{\rm B}N_{0}\gamma_{0}^{3}.
\label{equ:lumsypeak}
\end{equation}

For the jet opening angle $\theta_{j}$, the causality condition requires $\theta_{j}\Gamma\lesssim1$ \citep{2013A&A...558A.144C}, which is also supported by numerical simulations of axissymmetric, magnetically driven outflows \citep{2009MNRAS.394.1182K}, while constraints imposed by the SSC process \citep{2014ApJ...789..161N} and observations of radio cores \citep[e.g.,][]{2005AJ....130.1418J, 2009A&A...507L..33P} indicate $\theta_{j}\Gamma\gtrsim0.1-0.7$. In the calculations presented, we adopted a value of $\theta_{j}\Gamma=1$, and in this case, we have $\delta\lesssim\Gamma$ (with a viewing angle of roughly $\theta\gtrsim\theta_{j}$), of which the exact values will not significantly affect our conclusions.

The SSC emission may account for the emission at the peak of the IC component of BL Lacs, while the EC emission may account for that of FSRQs. The emissions at the IC peak can be at the Thomson or Klein-Nishina (KN) regimes. We will discuss these cases separately. The variability timescale can set an upper limit on the size of the emission region due to the causality, $R\lesssim\delta c\Delta t/(1+z)$. The variability timescales are not required to be the same in various sources \citep[see][for a review; and e.g. \citet{2011MNRAS.410..368B} and \citet{2011ApJ...736..131A} for the well-studied blazars 3C 454.3 and Mrk 421; and \citet{2013MNRAS.430.1324N} for a systematic study indicating a typical variability timescale in the source frame in the {\it Fermi}/LAT band of $\thickapprox$1 day; see also \citet{2014MNRAS.443.2940H}]{1997ARA&A..35..445U}, while for simplicity, the average value $\Delta t/(1+z)\thickapprox1$ day (in the source frame) is assumed in our calculation \citep[e.g.,][]{2014ApJS..215....5K, 2015ApJ...807...51Z}.

\subsection{SSC at the Thomson regime}

In the case of Thomson scattering, the SSC peak frequency follows
\begin{equation}
\frac{\nu'_{ssc}}{\nu'_{sy}}=\frac{4}{3}\gamma_{0}^{2}.
\label{equ:ssc_nu_r_th}
\end{equation}
Integrating the monochromatic luminosity (see Equations \ref{equ:jem_delta} and \ref{equ:Ngamma}), one can get the total synchrotron luminosity,
\begin{equation}
L'_{sy}=\nu'^{p}_{sy}L_{sy}'(\nu'^{p}_{sy})\sqrt{\frac{\pi\ln10}{b/4}}.
\label{equ:lumsypeak}
\end{equation}
Within the one-zone model, the average synchrotron energy density is \citep[see][]{2017ApJ...842..129C},
\begin{equation}
U_{sy-Th}=\frac{9}{4}\frac{L'_{sy}}{4\pi R^{2}c}.
\label{equ:usy_th}
\end{equation}
Therefore, the SSC peak luminosity follows
\begin{equation}
\frac{\nu'^{p}_{ssc}L'(\nu'^{p}_{ssc})}{\nu'^{p}_{sy}L'(\nu'^{p}_{sy})}
=\frac{U_{sy-Th}}{U_{B}}=\frac{\sigma_{T}}{2}\sqrt{\frac{\pi\ln10}{b/4}}RN_{0}\gamma_{0}^{3}.
\label{equ:ssc_lum_r_th}
\end{equation}

Combining the Equations \ref{equ:nu_sy}, \ref{equ:lumsypeak}, \ref{equ:ssc_nu_r_th}, \ref{equ:ssc_lum_r_th}, and
\begin{equation}
\left\{
\begin{array}{ll}
\nu=\delta\nu'\\
\nu L(\nu)=\delta^{4}\nu' L'(\nu')\\
R=c\Delta t\delta/(1+z)
\end{array}
\right.
\label{equ:tranR}
\end{equation}
we have
\begin{equation}
\left\{
\begin{array}{ll}
\gamma_{0}=8660.3\left(\frac{\nu_{ssc}^{p}}{10^{23}Hz}\frac{10^{15}Hz}{\nu_{sy}^{p}}\right)^{1/2}\\
\delta=9.0365\left(\frac{\nu^{p}_{sy}L(\nu^{p}_{sy})}{10^{45}erg/s}\right)^{1/4}
\left(\frac{\nu^{p}_{sy}}{10^{15}Hz}\right)^{-1/2}
\left(\frac{\nu^{p}_{ssc}L(\nu^{p}_{ssc})}{\nu^{p}_{sy}L(\nu^{p}_{sy})}\right)^{-1/4}
\left(\frac{\nu_{ssc}^{p}}{10^{23}Hz}\frac{10^{15}Hz}{\nu_{sy}^{p}}\right)^{1/2}
\left(\frac{\Delta t(1+z)}{1day}\right)^{-1/2}
\left(\frac{b}{2}\right)^{-1/8}\\
B=0.39533\left(\frac{\nu^{p}_{sy}L(\nu^{p}_{sy})}{10^{45}erg/s}\right)^{-1/4}
\left(\frac{\nu^{p}_{sy}}{10^{15}Hz}\right)^{3/2}
\left(\frac{\nu^{p}_{ssc}L(\nu^{p}_{ssc})}{\nu^{p}_{sy}L(\nu^{p}_{sy})}\right)^{1/4}
\left(\frac{\nu_{ssc}^{p}}{10^{23}Hz}\frac{10^{15}Hz}{\nu_{sy}^{p}}\right)^{-3/2}
\left(\frac{\Delta t(1+z)}{1day}\right)^{1/2}
\left(\frac{b}{2}\right)^{1/8}\\
R=2.3406\times10^{16}\left(\frac{\nu^{p}_{sy}L(\nu^{p}_{sy})}{10^{45}erg/s}\right)^{1/4}
\left(\frac{\nu^{p}_{sy}}{10^{15}Hz}\right)^{-1/2}
\left(\frac{\nu^{p}_{ssc}L(\nu^{p}_{ssc})}{\nu^{p}_{sy}L(\nu^{p}_{sy})}\right)^{-1/4}
\left(\frac{\nu_{ssc}^{p}}{10^{23}Hz}\frac{10^{15}Hz}{\nu_{sy}^{p}}\right)^{1/2}
\left(\frac{\Delta t}{1day(1+z)^{3}}\right)^{1/2}
\left(\frac{b}{2}\right)^{-1/8}\\
N_{0}=5.1991\times10^{-5}
\left(\frac{\nu^{p}_{sy}L(\nu^{p}_{sy})}{10^{45}erg/s}\right)^{-1/4}
\left(\frac{\nu^{p}_{sy}}{10^{15}Hz}\right)^{1/2}
\left(\frac{\nu^{p}_{ssc}L(\nu^{p}_{ssc})}{\nu^{p}_{sy}L(\nu^{p}_{sy})}\right)^{5/4}
\left(\frac{\nu_{ssc}^{p}}{10^{23}Hz}\frac{10^{15}Hz}{\nu_{sy}^{p}}\right)^{-2}
\left(\frac{\Delta t(1+z)}{1day}\right)^{-1/2}
\left(\frac{b}{2}\right)^{5/8}
\end{array}
\right.
\label{equ:jet_par_ssc_th}
\end{equation}

\subsection{SSC at the KN regime}
In the case of KN scattering ($\gamma_{0}h\nu'^{p}_{sy}\gtrsim m_{e}c^{2}$), the SSC peak frequency follows
\begin{equation}
\nu'^{p}_{ssc}\approx\frac{2}{\sqrt{3}}\frac{\gamma_{0}m_{e}c^{2}}{h}.
\label{equ:ssc_nu_kn}
\end{equation}
The effective SSC seed photon energy density can be easily derived through integrating from the minimum frequency to the critical frequency $\nu_{sy-c}'=(\sqrt{3}/2)(m_{e}c^{2}/h\gamma_{0})$,
\begin{equation}
U_{sy-KN}=U_{sy-Th}\frac{1}{\sqrt{\pi}}\int_{t_{c}}^{+\infty}e^{-t^{2}}dt
\label{equ:usy_kn}
\end{equation}
where
\begin{equation}
t_{c}=\sqrt{\frac{b}{4\ln10}}\ln\left(\frac{h\nu'^{p}_{sy}}{m_{e}c^{2}}\frac{h\nu'^{p}_{ssc}}{m_{e}c^{2}}\right)
\label{equ:tc_ssc_kn}
\end{equation}
In the case of the KN regime, the SSC peak luminosity follows
\begin{equation}
\frac{\nu'^{p}_{ssc}L'(\nu'^{p}_{ssc})}{\nu'^{p}_{sy}L'(\nu'^{p}_{sy})}
=\frac{U_{sy-KN}}{U_{B}}=\frac{\sigma_{T}}{2}\sqrt{\frac{\pi\ln10}{b/4}}RN_{0}\gamma_{0}^{3}
\frac{1}{\sqrt{\pi}}\int_{t_{c}}^{+\infty}e^{-t^{2}}dt.
\label{equ:ssc_lum_r_kn}
\end{equation}

Because of that the Equation \ref{equ:ssc_lum_r_kn} is actually an integral equation ($t_{c}$ is a function of $\delta$ through $\nu'^{p}_{sy,ssc}=\nu^{p}_{sy,ssc}/\delta$), the final estimated jet parameters cannot be analytically expressed as that in the Equation \ref{equ:jet_par_ssc_th}. The parameters ($\gamma_{0}$, $N_{0}$, $B$, $\delta$, and $R$) can be derived through numerically solving the set of Equations \ref{equ:nu_sy}, \ref{equ:lumsypeak}, \ref{equ:tranR}, \ref{equ:ssc_nu_kn}, and \ref{equ:ssc_lum_r_kn}.

\subsection{EC at the Thomson regime}

The FSRQs usually have very strong emission from the BLR and dust torus, and these photons can be IC scattered by relativistic electrons in the jet. In the frame of the jet, the external photon energy density will be enhanced and the frequency will be amplified: $U_{ext-Th}'\approx (17/12)\Gamma^{2}U_{ext}$ and $\nu_{ext}'\approx\Gamma\nu_{ext}$.
In this case, the EC peak luminosity follows
\begin{equation}
\frac{\nu'^{p}_{ec}L'(\nu'^{p}_{ec})}{\nu'^{p}_{sy}L'(\nu'^{p}_{sy})}
=\frac{U_{ext-Th}'}{U_{B}}\approx\frac{U_{ext}}{U_{B}}\frac{17}{12}\Gamma^{2}.
\label{equ:ec_lum_r_th1}
\end{equation}
and the EC peak frequency,
\begin{equation}
\nu'^{p}_{ec}\approx\frac{4}{3}\gamma_{0}^{2}\Gamma\nu_{ext}.
\label{equ:ec_nu_th}
\end{equation}
Combining the Equations \ref{equ:nu_sy}, \ref{equ:tranR}, \ref{equ:ec_lum_r_th1}, and \ref{equ:ec_nu_th}, we have \citep[see also][]{2011ApJ...735..108C}
\begin{equation}
\frac{\nu^{p}_{ec}L(\nu^{p}_{ec})}{\nu^{p}_{sy}L(\nu^{p}_{sy})}
=\frac{17e^{2}}{6\pi m_{e}^{2}c^{2}}\frac{U_{ext}}{\nu_{ext}^{2}}
\left(\frac{\nu^{p}_{ec}}{\nu^{p}_{sy}}\right)^{2}
\label{equ:ec_lum_r_th2}
\end{equation}
This means that the external photon properties (i.e., $U_{ext}/\nu_{ext}^{2}$) will be determined when given the synchrotron/EC peak frequency and luminosity. For the BLR, the emission is mainly contributed by Ly$\alpha$ lines, $\nu_{\rm ext-BLR}\approx2\times10^{15}$ Hz \citep{2009MNRAS.397..985G}. The dust torus reprocesses the disk emission into the IR band. As indicated by {\it Spitzer} observations, the typical peak frequency of IR dust torus emission is around $\nu_{\rm ext-IR}\approx3\times10^{13}$ \citep{2007ApJ...660..117C, 2009MNRAS.397..985G, 2013ApJ...773..176G}. Following \citet{2009MNRAS.397..985G}, the radiation from the reprocessed dust torus (IR) or BLR is described as a blackbody spectrum. The issue of whether the radiation regions are inside or outside the BLRs is highly debated. Considering the $\gamma$-ray photons can be absorbed via photon-photon pair production above $\gtrsim10$ GeV of BLR emission \citep[e.g.,][]{2006ApJ...653.1089L, 2009ApJ...699.2002B}, some authors suggested that the emission region should be outside the BLR \citep[e.g.,][]{2009ApJ...704...38S, 2009MNRAS.392L..40T, 2013MNRAS.435L..24T}. The broadband SED modeling of a large sample also suggests that the emission region may be outside the BLR for most sources \citep[e.g.,][]{2014ApJS..215....5K}. In this case, the IC/dust process will dominate the $\gamma$-ray emission \citep{2009MNRAS.397..985G}. In this paper, the IC/dust torus models are considered in our calculation.

We assume a the Doppler factor of $\delta=1/\Gamma\left(1-\beta\cos\theta\right)\approx\Gamma$ for the relativistic jet close to the line of sight in blazars with a viewing angle $\theta\lesssim1/\Gamma$ \citep{2005AJ....130.1418J, 2009A&A...507L..33P}. Combining the Equations \ref{equ:nu_sy}, \ref{equ:lumsypeak}, \ref{equ:ssc_lum_r_th}, \ref{equ:tranR}, \ref{equ:ec_lum_r_th1} (or the Equation \ref{equ:ec_lum_r_th2}), and \ref{equ:ec_nu_th}, the jet parameters can be derived (the SSC emission in this case is within the Thomson regime),
\begin{equation}
\left\{
\begin{array}{ll}
\gamma_{0}=
1641.5\left(\frac{\nu_{sy}^{p}}{10^{14}Hz}\right)^{1/2}
\left(\frac{\nu_{ec}^{p}}{10^{22}Hz}\frac{10^{14}Hz}{\nu_{sy}^{p}}\right)^{1/4}
\left(\frac{\nu^{p}_{sy}L(\nu^{p}_{sy})}{10^{46}erg/s}\right)^{-1/8}
\left(\frac{\nu^{p}_{ssc}L(\nu^{p}_{ssc})}{\nu^{p}_{sy}L(\nu^{p}_{sy})}\right)^{1/8}
\left(\frac{\nu_{ext}}{3\times10^{13}Hz}\right)^{-1/4}
\left(\frac{\Delta t(1+z)}{1day}\right)^{1/4}
\left(\frac{b}{2}\right)^{1/16}\\
\delta=9.6321
\left(\frac{\nu_{ec}^{p}}{10^{22}Hz}\frac{10^{14}Hz}{\nu_{sy}^{p}}\right)^{1/4}
\left(\frac{\nu^{p}_{sy}L(\nu^{p}_{sy})}{10^{46}erg/s}\right)^{1/8}
\left(\frac{\nu^{p}_{ssc}L(\nu^{p}_{ssc})}{\nu^{p}_{sy}L(\nu^{p}_{sy})}\right)^{-1/8}
\left(\frac{\nu_{ext}}{3\times10^{13}Hz}\right)^{-1/4}
\left(\frac{\Delta t(1+z)}{1day}\right)^{-1/4}
\left(\frac{b}{2}\right)^{-1/16}\\
B=1.0323
\left(\frac{\nu_{ec}^{p}}{10^{22}Hz}\frac{10^{14}Hz}{\nu_{sy}^{p}}\right)^{-3/4}
\left(\frac{\nu^{p}_{sy}L(\nu^{p}_{sy})}{10^{46}erg/s}\right)^{1/8}
\left(\frac{\nu^{p}_{ssc}L(\nu^{p}_{ssc})}{\nu^{p}_{sy}L(\nu^{p}_{sy})}\right)^{-1/8}
\left(\frac{\nu_{ext}}{3\times10^{13}Hz}\right)^{3/4}
\left(\frac{\Delta t(1+z)}{1day}\right)^{-1/4}
\left(\frac{b}{2}\right)^{-1/16}\\
R=2.4949\times10^{16}
\left(\frac{\nu_{ec}^{p}}{10^{22}Hz}\frac{10^{14}Hz}{\nu_{sy}^{p}}\right)^{1/4}
\left(\frac{\nu^{p}_{sy}L(\nu^{p}_{sy})}{10^{46}erg/s}\right)^{1/8}
\left(\frac{\nu^{p}_{ssc}L(\nu^{p}_{ssc})}{\nu^{p}_{sy}L(\nu^{p}_{sy})}\right)^{-1/8}
\left(\frac{\nu_{ext}}{3\times10^{13}Hz}\right)^{-1/4}
\left(\frac{\Delta t(1+z)}{1day}\right)^{3/4}
\left(\frac{b}{2}\right)^{-1/16}\\
N_{0}=7.1622\times10^{-3}\left(\frac{\nu_{sy}^{p}}{10^{14}Hz}\right)^{-3/2}
\left(\frac{\nu_{ec}^{p}}{10^{22}Hz}\frac{10^{14}Hz}{\nu_{sy}^{p}}\right)^{-1}
\left(\frac{\nu^{p}_{sy}L(\nu^{p}_{sy})}{10^{46}erg/s}\right)^{1/4}
\left(\frac{\nu^{p}_{ssc}L(\nu^{p}_{ssc})}{\nu^{p}_{sy}L(\nu^{p}_{sy})}\right)^{3/4}
\left(\frac{\nu_{ext}}{3\times10^{13}Hz}\right)
\left(\frac{\Delta t(1+z)}{1day}\right)^{-3/2}
\left(\frac{b}{2}\right)^{3/8}\\
U_{ext}=3.2260\times10^{-4}\left(\frac{\nu^{p}_{ec}L(\nu^{p}_{ec})}{\nu^{p}_{sy}L(\nu^{p}_{sy})}\right)
\left(\frac{\nu_{ec}^{p}}{10^{22}Hz}\frac{10^{14}Hz}{\nu_{sy}^{p}}\right)^{-2}
\left(\frac{\nu_{ext}}{3\times10^{13}Hz}\right)^{2}
\end{array}
\right.
\label{equ:jet_par_ec_th}
\end{equation}

\subsection{EC at the KN regime}

In the case of the KN regime ($\gamma_{0}\Gamma h\nu_{\rm ext}\gtrsim m_{e}c^{2}$), the EC peak frequency is
\begin{equation}
\nu'^{p}_{ec}\approx\frac{2}{\sqrt{3}}\frac{\gamma_{0}m_{e}c^{2}}{h}.
\label{equ:ec_nu_kn}
\end{equation}
The effective EC seed photon energy density can be easily calculated through integrating from the minimum frequency to the critical frequency $\nu_{ext-c}'=(\sqrt{3}/2)(m_{e}c^{2}/h\gamma_{0})$,
\begin{equation}
U_{ext-KN}'=U_{ext-Th}'\frac{15}{\pi^{4}}\int_{0}^{x_{c}}\frac{x^{3}}{e^{x}-1}dx,
\label{equ:uext_kn}
\end{equation}
where\footnote{The factor 2.82 accounts for the difference between the peak frequency of the blackbody spectrum and the $kT/h$.},
\begin{equation}
x_{c}=2.82\frac{m_{e}^{2}c^{2}}{h\nu_{ext}}\frac{m_{e}^{2}c^{2}}{h\nu_{ec}}.
\label{equ:xc_ec_kn}
\end{equation}
In the case of the KN regime, the EC peak luminosity follows
\begin{equation}
\frac{\nu'^{p}_{ec}L'(\nu'^{p}_{ec})}{\nu'^{p}_{sy}L'(\nu'^{p}_{sy})}
=\frac{U_{ext-KN}'}{U_{B}}
=\frac{U_{ext}}{U_{B}}\frac{17}{12}\Gamma^{2}\int_{0}^{x_{c}}\frac{x^{3}}{e^{x}-1}dx.
\label{equ:ec_lum_r_kn}
\end{equation}

Combining the Equations \ref{equ:nu_sy}, \ref{equ:lumsypeak}, \ref{equ:ssc_lum_r_th}, \ref{equ:tranR}, and \ref{equ:ec_nu_kn}, we have (the SSC in this case is within the Thomson regime)
\begin{equation}
\left\{
\begin{array}{ll}
\gamma_{0}=61460
\left(\frac{\nu^{p}_{sy}L(\nu^{p}_{sy})}{10^{46}erg/s}\right)^{-1/8}
\left(\frac{\nu^{p}_{ssc}L(\nu^{p}_{ssc})}{\nu^{p}_{sy}L(\nu^{p}_{sy})}\right)^{1/8}
\left(\frac{\nu_{sy}^{p}}{10^{17}Hz}\right)^{1/4}
\left(\frac{\nu_{ec}^{p}}{10^{26}Hz}\frac{10^{17}Hz}{\nu_{sy}^{p}}\right)^{1/2}
\left(\frac{\Delta t(1+z)}{1day}\right)^{1/4}
\left(\frac{b}{2}\right)^{1/16}\\
\delta=11.404
\left(\frac{\nu^{p}_{sy}L(\nu^{p}_{sy})}{10^{46}erg/s}\right)^{1/8}
\left(\frac{\nu^{p}_{ssc}L(\nu^{p}_{ssc})}{\nu^{p}_{sy}L(\nu^{p}_{sy})}\right)^{-1/8}
\left(\frac{\nu_{sy}^{p}}{10^{17}Hz}\right)^{-1/4}
\left(\frac{\nu_{ec}^{p}}{10^{26}Hz}\frac{10^{17}Hz}{\nu_{sy}^{p}}\right)^{1/2}
\left(\frac{\Delta t(1+z)}{1day}\right)^{-1/4}
\left(\frac{b}{2}\right)^{-1/16}\\
B=0.62198
\left(\frac{\nu^{p}_{sy}L(\nu^{p}_{sy})}{10^{46}erg/s}\right)^{1/8}
\left(\frac{\nu^{p}_{ssc}L(\nu^{p}_{ssc})}{\nu^{p}_{sy}L(\nu^{p}_{sy})}\right)^{-1/8}
\left(\frac{\nu_{sy}^{p}}{10^{17}Hz}\right)^{3/4}
\left(\frac{\nu_{ec}^{p}}{10^{26}Hz}\frac{10^{17}Hz}{\nu_{sy}^{p}}\right)^{-3/2}
\left(\frac{\Delta t(1+z)}{1day}\right)^{-1/4}
\left(\frac{b}{2}\right)^{-1/16}\\
R=2.9539\times10^{16}
\left(\frac{\nu^{p}_{sy}L(\nu^{p}_{sy})}{10^{46}erg/s}\right)^{1/8}
\left(\frac{\nu^{p}_{ssc}L(\nu^{p}_{ssc})}{\nu^{p}_{sy}L(\nu^{p}_{sy})}\right)^{-1/8}
\left(\frac{\nu_{sy}^{p}}{10^{17}Hz}\right)^{-1/4}
\left(\frac{\nu_{ec}^{p}}{10^{26}Hz}\frac{10^{17}Hz}{\nu_{sy}^{p}}\right)^{1/2}
\left(\frac{\Delta t(1+z)}{1day}\right)^{3/4}
\left(\frac{b}{2}\right)^{-1/16}\\
N_{0}=1.1526\times10^{-7}
\left(\frac{\nu^{p}_{sy}L(\nu^{p}_{sy})}{10^{46}erg/s}\right)^{1/4}
\left(\frac{\nu^{p}_{ssc}L(\nu^{p}_{ssc})}{\nu^{p}_{sy}L(\nu^{p}_{sy})}\right)^{3/4}
\left(\frac{\nu_{sy}^{p}}{10^{17}Hz}\right)^{-1/2}
\left(\frac{\nu_{ec}^{p}}{10^{26}Hz}\frac{10^{17}Hz}{\nu_{sy}^{p}}\right)^{-2}
\left(\frac{\Delta t(1+z)}{1day}\right)^{-3/2}
\left(\frac{b}{2}\right)^{3/8}
\end{array}
\right.
\label{equ:jet_par_ec_kn}
\end{equation}
After getting these parameters, the external energy density, $U_{\rm ext}$, can be easily derived through combining the Equations \ref{equ:tranR} and \ref{equ:ec_lum_r_kn}.

As discussed above, we use the SSC model for all BL Lacs and the EC model for all FSRQs. The input parameters include $\nu_{\rm sy}^{\rm p}$, $\nu_{\rm ssc/ec}^{\rm p}$, $\nu^{p}_{sy}L(\nu^{p}_{sy})$, $\nu^{p}_{ssc}L(\nu^{p}_{ssc})$ and $P_{1}$. The criterion from the Thomson to KN regimes in the SSC model is $\gamma_{0}h\nu'^{p}_{sy}\gtrsim m_{e}c^2$ (see Section 3.2), and that in the EC model is $\gamma_{0}\Gamma h\nu_{ext}\gtrsim m_{e}c^{2}$ (see Section 3.4).

\section{Results and Discussion}
\label{sec:resultdiscussion}

the {\it COS B} satellite first detected 3C 273 as a $\gamma$-ray blazar \citep{1978Natur.275..298S}. After that, nearly 100 blazars were discovered by the Energetic Gamma-Ray Experiment Telescope (EGRET) onboard the {\it Compton Gamma Ray Observatory} \citep{1999ApJS..123...79H, 2007ApJ...657..706N}. The 20-fold (sensitivity) improvement of {\it Fermi}/LAT has now detected more than 1000 blazars, which allows us to do population studies. For all 1392 $\gamma$-ray blazars in our sample, we estimate the jet physical parameters.

\begin{figure*}
\centering
\includegraphics[height=5.7cm]{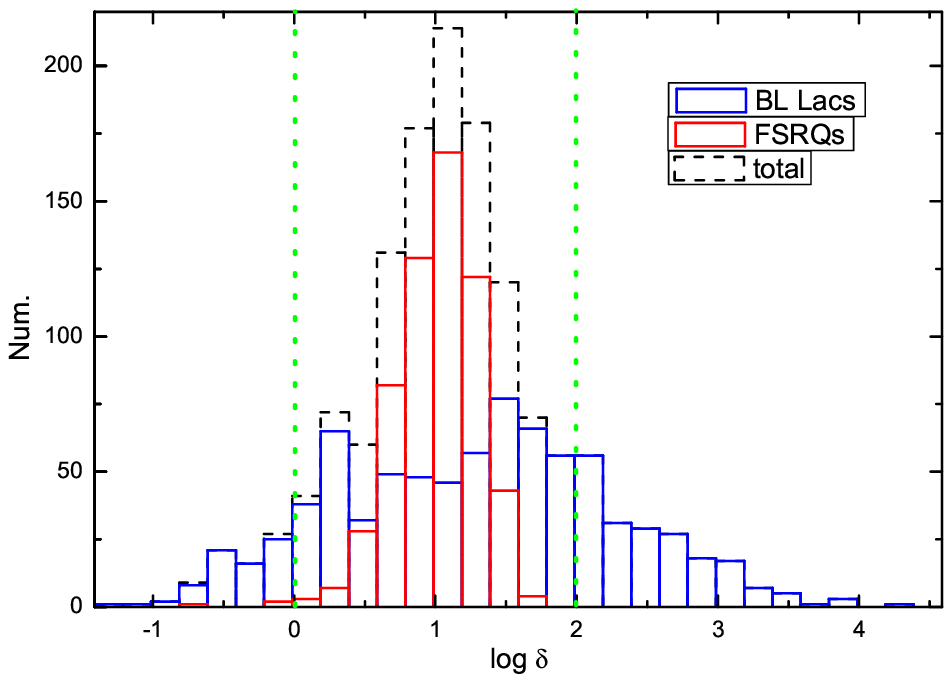}
\includegraphics[height=5.7cm]{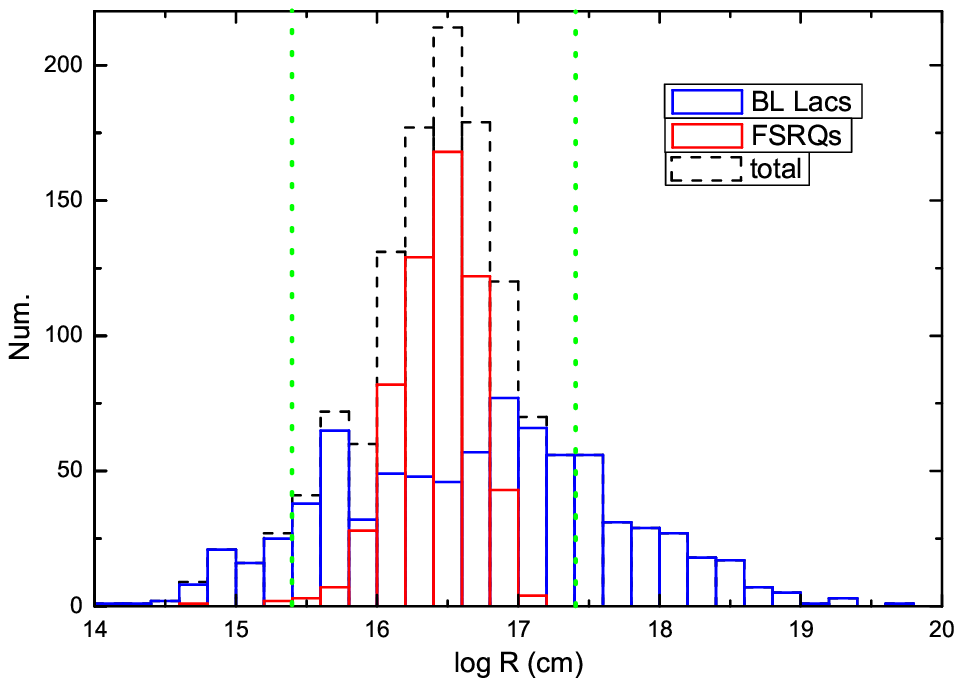}
\caption{Distributions of $\delta$ and $R$. Left panel: The median values of the Doppler factor $\delta$ of FSRQs, BL Lacs, and total blazars are 10.7, 22.4, and 13.1, respectively. For sources with $1<\delta<100$ (i.e., within the two green dotted lines), the median values of $\delta$ of FSRQs and BL Lacs are 10.7 and 14.3, respectively. Right panel: The median values of the size of the emission region $R$ of FSRQs, BL Lacs, and total blazars are 2.76, 5.79, and 3.38 $\times10^{16}$ cm, respectively. For sources with $1<\delta<100$ (i.e., within the two green dotted lines), the median values of $R$ of FSRQs and BL Lacs are 2.78 and 3.70 $\times10^{16}$ cm, respectively. See also the Table \ref{tab:median}.}
\label{fig:deltaR}
\end{figure*}

\begin{figure*}
\centering
\includegraphics[height=5.7cm]{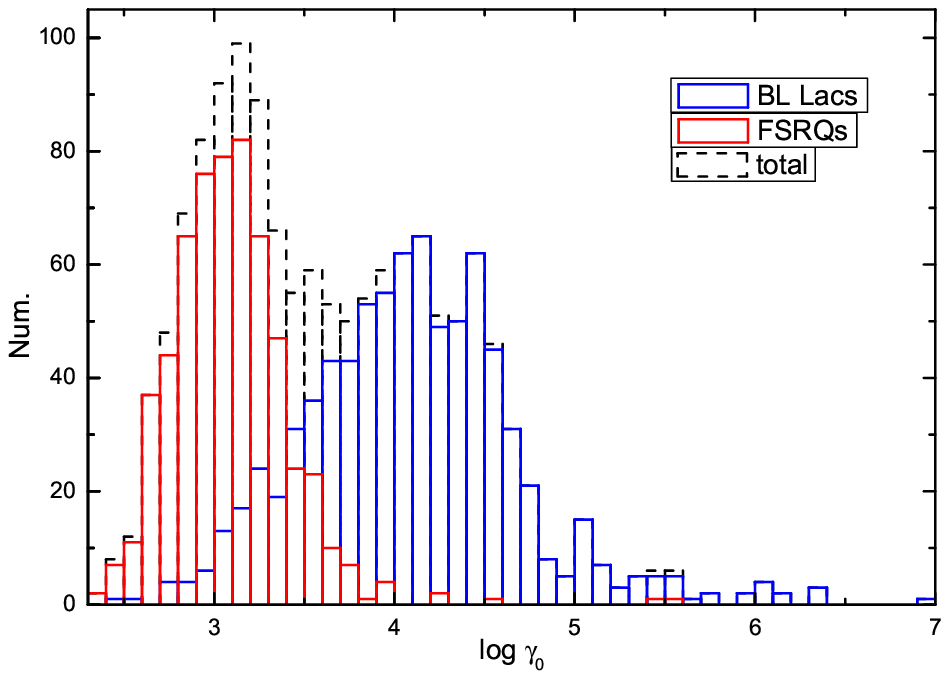}
\includegraphics[height=5.7cm]{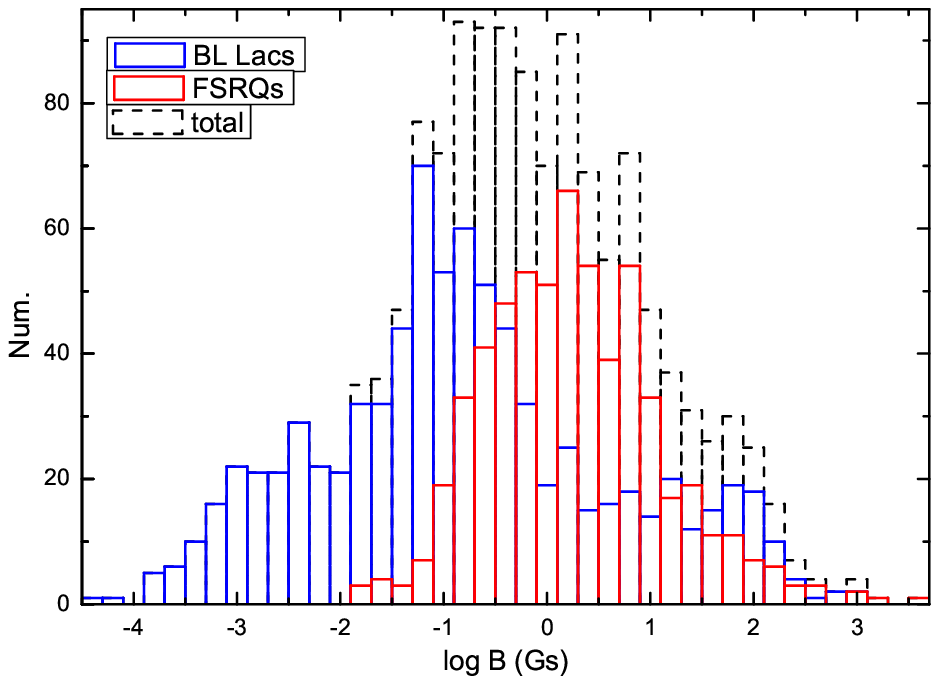}
\includegraphics[height=5.7cm]{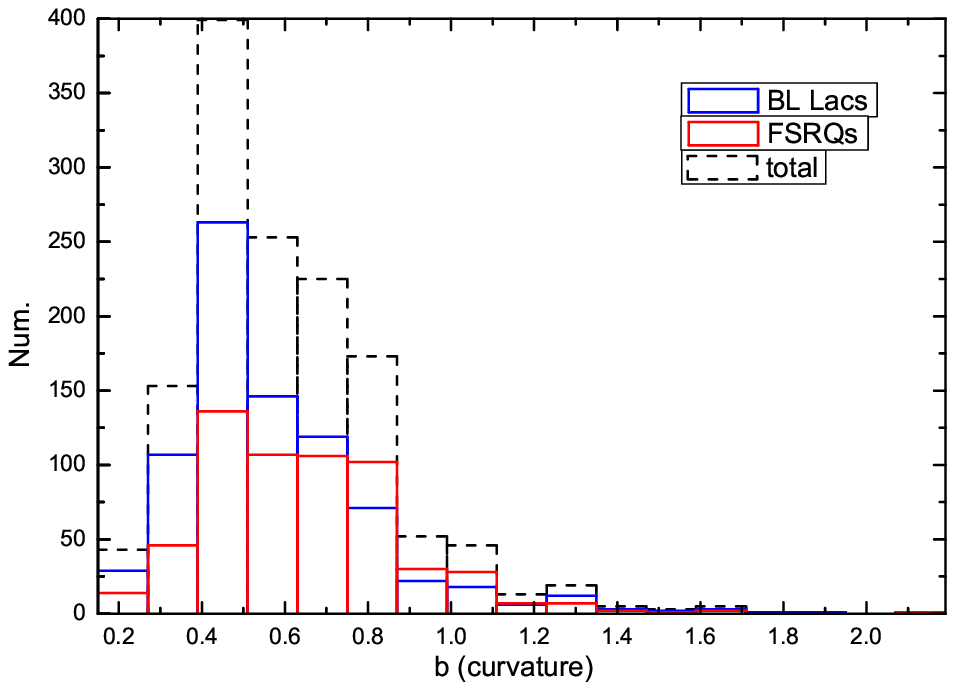}
\includegraphics[height=5.7cm]{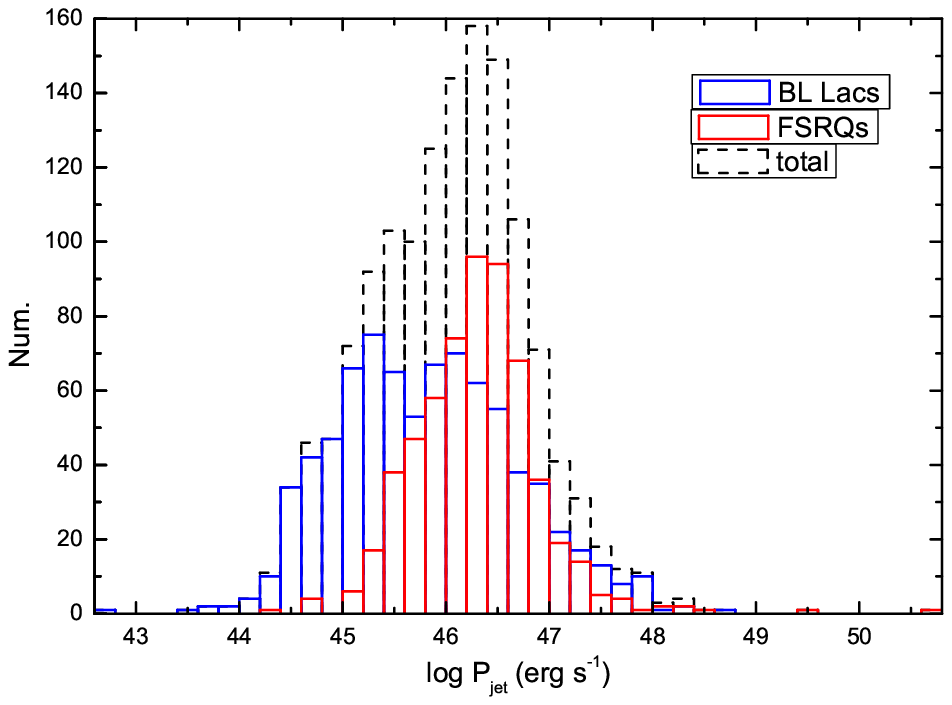}
\caption{Distributions of some jet parameters. Upper left panel: The median values of the electron peak energy $\gamma_{0}$ of FSRQs, BL Lacs, and total blazars are 1167.8, 12077, and 3646.1, respectively. Upper right panel: The median values of the strength of the magnetic field $B$ of FSRQs, BL Lacs, and total blazars are 1.56, 0.119, and 0.446, respectively. Bottom left panel: The median values of the curvature $b$ of FSRQs, BL Lacs, and total blazars are 0.6, 0.55, and 0.55, respectively. Bottom right panel: The median values of the jet power $P_{\rm jet}$ of FSRQs, BL Lacs, and total blazars are 20.0, 6.3, and 12.0 $\times10^{45}$ erg s$^{-1}$, respectively. See also the Table \ref{tab:median}.}
\label{fig:distributions}
\end{figure*}

We firstly calculate the Doppler factor ($\delta$) and the size of the emission region ($R$), for which (histogram) distributions are presented in Figure \ref{fig:deltaR}. It can be seen that $R$ and $\delta$ follow the same distributions, which is due to the assumptions $R\lesssim\delta c\Delta t/(1+z)$ and $\Delta t/(1+z)\approx1$ day in our calculation. The median values of the Doppler factors of FSRQs, BL Lacs, and total blazars are $\delta\simeq$ 10.7, 22.3, and 13.1, respectively. The significantly larger of $\delta$ of BL Lacs relative to that of FSRQs seems to be inconsistent with estimations from other methods. The independent methods found that the Doppler factors of BL Lacs are comparable with or even smaller than those of FSRQs, such as from radio variability \citep{1999ApJ...521..493L, 2009PASJ...61..639F, 2010A&A...512A..24S, 2017MNRAS.466.4625L}, SED modeling \citep{1998MNRAS.301..451G, 1993ApJ...407...65G}, pair production \citep{1993ApJ...410..609M, 2014RAA....14.1135F}, or apparent superluminal motions \citep{2005AJ....130.1418J, 2009A&A...494..527H}. From Figure \ref{fig:deltaR}, we find that some BL Lacs present extremely large values of $\delta$, even reaching $\sim10^{4}$, which values may be incorrect for blazars. The typical values of the Doppler factor in blazars range from a few to $\sim50$ based on various methods as mentioned above \citep[e.g.,][]{1999ApJ...521..493L, 2009PASJ...61..639F, 2010A&A...512A..24S, 2017MNRAS.466.4625L, 1998MNRAS.301..451G, 1993ApJ...407...65G, 1993ApJ...410..609M, 2014RAA....14.1135F, 2005AJ....130.1418J, 2009A&A...494..527H, 2014Natur.515..376G, 2015MNRAS.448.1060G}. There is also a small fraction of BL Lacs having very low values of Doppler factor ($\delta<1$). In our model, the Doppler factor is assumed to be equal to the jet bulk Lorentz factor $\delta=\Gamma$, which must be larger than 1. For these sources with extreme Doppler factors, the estimations of other jet parameters may also be incorrect. We calculate the median value of the Doppler factors of all BL Lacs with $1<\delta<100$, which gives $\delta_{\rm m}\simeq14.3$. This median value is used for 268 BL Lacs with extreme Doppler factors ($\delta<1$ or $\delta>100$) and determining other jet parameters. In this case, $\delta$ is a new input parameter other than $\nu_{\rm sy}^{\rm p}$, $\nu_{\rm ssc}^{\rm p}$, $\nu^{p}_{sy}L(\nu^{p}_{sy})$, $\nu^{p}_{ssc}L(\nu^{p}_{ssc})$, and $P_{1}$. Due to the large uncertainty of the estimation of $\nu_{\rm ssc}^{\rm p}$ \citep[see][and discussion below]{2010ApJ...716...30A}, we therefore use $\delta$ instead of $\nu_{\rm ssc}^{\rm p}$ as the input parameter to calculate other jet parameters. There are no FSRQs having $\delta>100$ and only three FSRQs having $\delta<1$. We use the same method to determine the other jet parameters of these three FSRQs (using $\delta_{\rm m}\simeq10.7$, the median value of $\delta$ of all FSRQs with $\delta>1$). We calculate all jet parameters and present them in Table \ref{tab:data}, where sources with $\delta<1$ or $\delta>100$ are labeled with an asterisk. The full version of the Table \ref{tab:data} is available in online ASCII form, which can be downloaded publicly. In the Figure \ref{fig:distributions}, we present the (histogram) distributions of some jet parameters, and the median values of these are listed in Table \ref{tab:median}.

\begin{figure}
\begin{center}
{\includegraphics[width=1.0\linewidth]{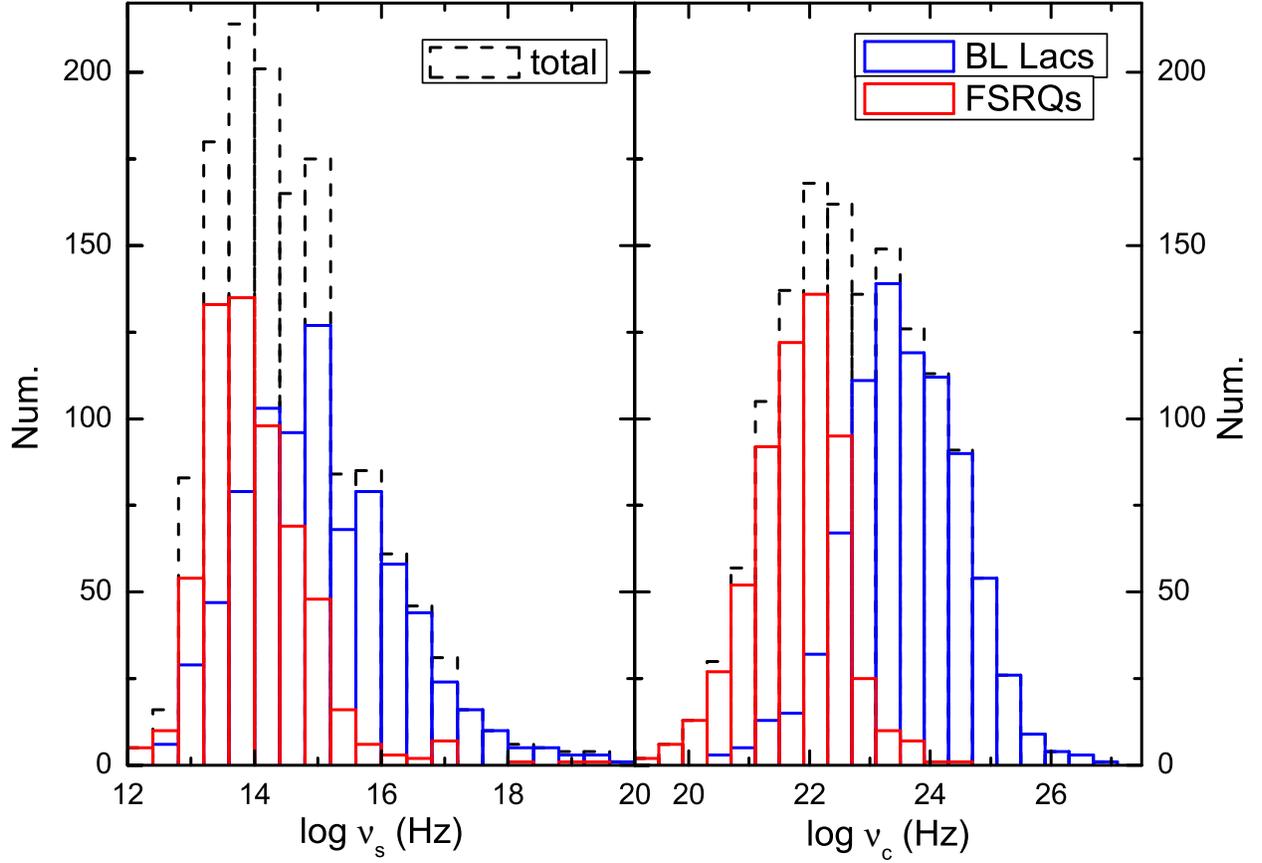}}
\end{center}
\caption{Distribution of synchrotron (left panel) and IC (right panel) peak frequency (measured in the AGN frame). The median values are $\nu_{\rm s}=$(0.689, 8.13, 2.48)$\times10^{14}$ Hz and $\nu_{\rm IC}=$(0.692, 33.1, 5.75)$\times10^{22}$ Hz for FSRQs, BL Lacs, and total blazars, respectively.}
\label{fig:nu_peak}
\end{figure}

The upper left panel in Figure \ref{fig:distributions} is the distribution of the electron peak energy $\gamma_{0}$. The red line refers to FSRQs, the blue line indicates BL Lacs, and the black dashed line is for the total blazars. It clearly shows two separated populations/dichotomic distributions for FSRQs and BL Lacs, with median values $\gamma_{0}\simeq$1167.8 and 12077, respectively, which are comparable to previous studies\footnote{The median value of $\gamma_{0}\simeq$1167.8 of FSRQs in this paper is slightly larger than the previous result, where $\gamma_{0}$ ranges from 100 to 1000 \citep[e.g.,][]{2015ApJ...807...51Z, 2015MNRAS.448.1060G}.} \citep[e.g.,][]{2015ApJ...807...51Z, 2015MNRAS.448.1060G, 2017arXiv171110625Q}. The median value for total blazars is $\gamma_{0}\simeq$3646.1. These separated populations between FSRQs and BL Lacs are actually results from the separated distribution of synchrotron peak frequency (also for the IC component) between FSRQs and BL Lacs, which are presented in Figure \ref{fig:nu_peak}. These separated populations may reflect that FSRQs and BL Lacs are different in nature. As discussed above, the high accretion rate makes FSRQs usually luminous AGNs having very strong line / dust torus emission. Therefore, nonthermal electrons in the jets of FSRQs will suffer efficient cooling and attain a smaller typical energy.  The low accretion rate makes BL Lacs low-luminosity AGNs, with no or very weak emission lines. Therefore, the electrons in BL Lac jets suffer less cooling, which leads to a larger typical energy. These results are consistent with the so-called blazar sequence \citep[see, e.g.,][]{1998MNRAS.299..433F, 1998MNRAS.301..451G, 2011ApJ...735..108C}. Blazars were classified as low-, intermediate-, and high-synchrotron-peaked (LSP, ISP, and HSP) blazars based on their synchrotron peak frequency by \citet{2010ApJ...715..429A} and later by \citet{2016ApJS..226...20F} based on their Bayesian classification of peak frequency for a larger sample of {\it Fermi}/LAT-detected blazars. From LSP to HSP blazars, the luminosity forms a gradual decreasing trend, which is called the blazar sequence \citep{1998MNRAS.299..433F, 1998MNRAS.301..451G, 2011ApJ...735..108C}. However, some authors claimed that this sequence was from a selection effect, because after being corrected by the Doppler-beaming effect, the negative correlation between the peak frequency and luminosity disappeared \citep[see][]{2008A&A...488..867N, 2009RAA.....9..168W, 2017ApJ...835L..38F}. The IC emission of smaller-energy electrons in FSRQs corresponds to softer $\gamma$-ray emission, while the higher-energy electrons in BL Lacs produce harder $\gamma$-ray emission. This is consistent with the fact that FSRQs and BL Lacs are distributed at two separate areas in the $L_{\gamma}-\Gamma$ plane, as shown in Figure \ref{fig:blazar_divide} \citep[see also][]{2009MNRAS.396L.105G, 2009ApJ...700..597A, 2015ApJ...810...14A}.

The distribution of the magnetic field is shown in the upper right panel of Figure \ref{fig:distributions}. The median values of FSRQs, BL Lacs, and total blazars are $B\simeq$ 1.56, 0.119, and 0.446 Gs, respectively. These values are roughly consistent with the blazar sample SED modeling results \citep{2015MNRAS.448.1060G}, in which the typical value of the magnetic field strength of FSRQs are found to range from 1 to 10 Gs and BL Lacs widely range from 0.001 to 10 Gs \citep[see, e.g.,][]{2010MNRAS.401.1570T, 2012ApJ...752..157Z, 2017arXiv171106282C}.

The distribution of the curvature $b$ of the electron energy distribution (see Equation \ref{equ:Ngamma}) is presented in the bottom left panel of Figure \ref{fig:distributions}. Note that we calculate this parameter $b=5P_{1}$, where $P_{1}$ is the curvature of the observed SED of the synchrotron component, whose values are given by \citet{2016ApJS..226...20F}. The median value of FSRQs ($b\simeq0.6$) is only slightly larger than that of BL Lacs ($b\simeq0.55$), with the median of total blazars $b\simeq0.55$ \citep[see][]{2014ApJ...788..179C}.

The uncertainty of the estimation of these parameters comes from the input parameters. An important contribution comes from the IC peak frequency $\nu_{IC}^{p}$, which is derived from the $\gamma$-ray photon index through $\log\nu_{IC}^{p}=-4\Gamma+31.6$ \citep[Equation 5 in][]{2010ApJ...716...30A}. As proposed by \citet{{2010ApJ...716...30A}}, the error of the estimated $\log\nu_{IC}^{p}$ is about $\sim0.7$, which leads to uncertainty of the estimated jet parameters. For the SSC model, we have a corresponding uncertainty about $\Delta\log\gamma_{0}\sim0.35$, $\Delta\log\delta\sim0.35$, $\Delta\log B\sim1.05$ and $\Delta\log R\sim0.35$ (simply for Thomson scatting; see Equation \ref{equ:jet_par_ssc_th}). For the EC model, we have a corresponding uncertainty about $\Delta\log\gamma_{0}\sim0.175$, $\Delta\log\delta\sim0.175$, $\Delta\log B\sim0.525$ and $\Delta\log R\sim0.175$ (simply for Thomson scatting; see Equation \ref{equ:jet_par_ec_th}).

\subsection{Disk-jet Connection and Jet Launching}
\label{sec:jet_disk}

The most promising scenario for launching astrophysical relativistic jets involves large-scale magnetic fields anchored in rapidly rotating compact objects. The idea of driving outflows by rotating magnetic fields, originally invented by \citet{1967ApJ...148..217W} to explain the spindown of young stars, was successfully applied to pulsar winds \citep{1969ApJ...158..727M} and became a dominant mechanism in theories of relativistic jets in AGNs \citep{1987ApJ...315..504L, 1992ApJ...394..459L, 2004ApJ...605..656V} and $\gamma$-ray bursts \citep[e.g.,][]{2001A&A...369..694S, 2001ApJ...563L.129V}. Powerful jets in AGNs can be powered by the innermost portions of accretion disks and/or by rapidly rotating BHs \citep{1977MNRAS.179..433B, 1982MNRAS.199..883B, 2010LNP...794..233S}. The correlation between the jet and the accretion disk/BLR has been presented in many studies \citep[e.g.,][]{1999MNRAS.307..802C, 2001ApJ...555..650H, 2003A&A...409..887W, 2014Natur.515..376G, 2014AJ....147...17L, 2016MNRAS.462.1542S, 2016NewA...46....9D}. Because the disk continuum and line emission are usually missing in BL Lacs, we study the jet-disk connection only for FSRQs through comparing the accretion (e.g., accretion disk/BLR/BH mass) with the jet properties.

The disk bolometric luminosity $L_{\rm bol}$ can be derived from a continuum luminosity based on the empirical relation \citep{2011ApJS..194...45S}. In order to avoid contamination by the nonthermal continuum, the accretion disk luminosity can be derived from the luminosity of broad emission lines as a proxy of $L_{\rm disk}\approx10L_{\rm BLR}$ \citep{2013MNRAS.431..210C}, and the total luminosity of broad emission lines can be reconstructed through visible broad lines \citep[see, e.g.,][]{1991ApJ...373..465F, 2001AJ....122..549V}. H$\beta$, Mg II, and C IV \citep{2012ApJ...748...49S} are widely used to scale to the quasar template spectrum of \citet{1991ApJ...373..465F} to calculate the total BLR luminosity $L_{\rm BLR}$. In \citet{1991ApJ...373..465F}, the Ly$\alpha$ is used as a reference of 100, and the total relative BLR flux is 555.77, of which H$\beta$ is 22, Mg II is 34 and CIV is 63 \citep[see also, e.g.,][]{2001AJ....122..549V, 2014Natur.515..376G}. The BH mass and accretion rate are fundamental parameters of AGNs. The virial method is now the most widely used to estimate the BH mass \citep[e.g.,][]{2001MNRAS.327.1111G, 2004A&A...424..793W, 2005ApJ...630..122G}. The uncertainty of a BH mass derived in this way is a factor of $\sim3-4$ \citep[][]{2006ApJ...641..689V}. Other methods include using the relation between BH mass and stellar velocity dispersion \citep[e.g.,][]{2002A&A...389..742W} and between BH mass and host-galaxy bulge luminosity \citep[e.g.,][]{2001ApJ...553..677L}. \citet{2012ApJ...748...49S} reported the optical spectroscopy of a large sample of $\gamma$-ray-detected blazars, and estimated their BH mass through the virial method. We collect their available data (BH mass and emission lines) for our sample and finally get 144 FSRQs. As in the method discussed above, we use their optical emission lines to scale to the quasar template spectrum to calculate the total BLR luminosity. Because of that the BLR luminosities derived from various emission lines are consistent with each other \citep[e.g.,][]{1991ApJ...373..465F, 2011ApJS..194...45S, 2012ApJ...748...49S}, we use the average value of the BLR luminosity if more than one emission line is available. The disk luminosity is then derived by $L_{\rm disk}\approx10L_{\rm BLR}$. \citet{2011ApJS..194...45S} studied the properties of quasars in the Sloan Digital Sky Survey Data Release 7 catalog. They calculated the virial BH mass and bolometric luminosity $L_{\rm bol}$ through the correlation between $L_{\rm bol}$ and continuum luminosity as presented in \citet{2006ApJS..166..470R}. The disk luminosity can be calculated $L_{\rm disk}=L_{\rm bol}/2$ \citep[see][]{2013MNRAS.431..210C}. We further compile available data from \citet{2011ApJS..194...45S}, and get another 46 FSRQs with disk luminosity and 44 FSRQs with BH mass. We note that some famous FSRQs are not included in these samples, e.g., 3C 273 and 3C 454.3. We then search the literature for FSRQs with well-measured disk luminosity and BH mass \citep{2014MNRAS.441.3375X}. Finally, we have 232 FSRQs with measured disk luminosity and 229 with measured BH mass, 222 FSRQs have both values. The 3 FSRQs with Doppler factors $\delta<1$ are not included in these sources.

\begin{figure}
\begin{center}
{\includegraphics[width=1.0\linewidth]{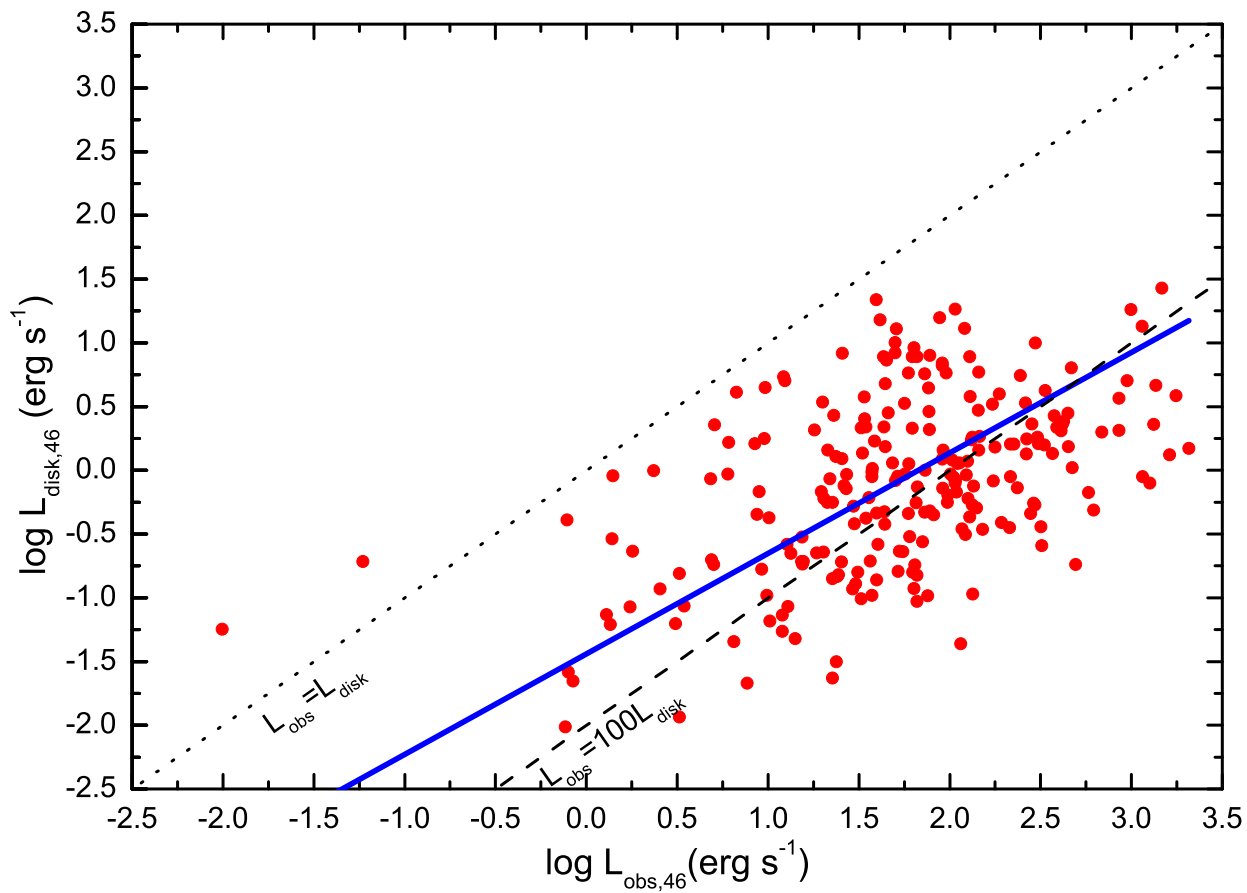}}
\end{center}
\caption{Jet bolometric luminosity versus disk luminosity. The solid blue line shows a robust correlation, $\log L_{\rm disk,46}=(0.788\pm0.076)\log L_{\rm obs,46}-(1.44\pm0.14)$, with a chance probability $p=4.76\times10^{-15}$ through a Pearson test. The black dotted line represents $L_{\rm obs}=L_{\rm disk}$ and the black dashed line $L_{\rm obs}=100L_{\rm disk}$.}
\label{fig:Lobs_Ldisk}
\end{figure}

Integrating the observed broadband SED, including both synchrotron and IC components, we can calculate the jet bolometric luminosity $L_{\rm obs}$. The Figure \ref{fig:Lobs_Ldisk} shows $L_{\rm obs}$ as a function of $L_{\rm disk}$ for the 232 FSRQs with well-measured disk luminosity. These show a robust correlation\footnote{$L_{\rm disk,46}=(L_{\rm disk}/10^{46})$ erg s$^{-1}$ and $L_{\rm obs,46}=(L_{\rm obs}/10^{46})$ erg s$^{-1}$.} $\log L_{\rm disk,46}=(0.788\pm0.076)\log L_{\rm obs,46}-(1.44\pm0.14)$ (solid blue line) with a chance probability $p=4.76\times10^{-15}$ (Pearson test). The black dotted line represents $L_{\rm obs}=L_{\rm disk}$ and the black dashed line $L_{\rm obs}=100L_{\rm disk}$, which indicates that the jet bolometric luminosity is of the order of $\sim$100 times the disk luminosity $L_{\rm obs}\approx100L_{\rm disk}$. The entire radiation power of the jet can be estimated through $P_{\rm r}=2L_{\rm obs}\Gamma^{2}/\delta^{4}\approx2L_{\rm obs}/\delta^{2}$, which is the power that the jet expends in producing the nonthermal radiation. Based on VLBI-detected superluminal motion, or broadband SED model, the jet Doppler factor is found to be an order of $\delta\gtrsim$10 \citep[e.g.,][]{2013AJ....146..120L, 2015MNRAS.448.1060G}. Our estimation gives a median value of $\delta=10.7$ for FSRQs (see the discussion above and/or the upper left panel of Figure \ref{fig:distributions}). In this case, we have the entire jet radiation power will be on the order of the disk luminosity $P_{\rm r}\approx L_{\rm disk}$. The jet radiative efficiency is believed to be on the order of $P_{\rm r}/P_{\rm jet}\sim$10\%, which holds for AGNs, $\gamma$-ray bursts, and even for BH X-ray binaries \citep{2012Sci...338.1445N, 2013ApJ...774L...5Z, 2014ApJ...780L..14M}, which gives an inevitable consequence that the jet power, $P_{\rm jet}\thickapprox10P_{\rm r}$, is larger than the disk luminosity, $P_{\rm jet}\approx10L_{\rm disk}$. This suggests that the jet-launching processes and the way of transporting energy from the vicinity of the BH to infinity must be very efficient.

The relativistic jet may be driven from the rapidly rotating BH. Increasing the spin of the BH shrinks the innermost stable orbit while increasing the accretion disk radiative efficiency $\eta_{\rm acc}=L_{\rm disk}/\dot{M}c^{2}$ to a maximum value $\eta_{\rm acc}\approx0.3$ \citep{1974ApJ...191..507T}, where $\dot{M}$ is the mass accretion rate. Assuming the accretion disk radiative efficiency $\eta_{\rm acc}=0.3$, we calculate the accretion power $L_{\rm acc}=\dot{M}c^{2}=L_{\rm disk}/\eta_{\rm acc}$ \citep[same as that in, e.g.,][]{2014Natur.515..376G}.

With the jet physical parameters, we calculate the jet power through\footnote{The factor of 2 accounts for the two jets.}, $P_{\rm jet}=2\pi R^{2}c\Gamma^{2}(U_{\rm B}+U_{\rm e}+U_{\rm p})$, where $U_{\rm e}$ and $U_{\rm p}$ are the electron and proton energy density in the jet comoving frame \citep[see, e.g.,][]{2008MNRAS.385..283C, 2014Natur.515..376G}. We adopt the conventional assumption that jet power is carried by electrons and protons (with one cold proton per emitting electron). In this case, $U_{\rm e}$ and $U_{\rm p}$ can be easily derived for the log-parabolic distribution of electrons (see Equation \ref{equ:Ngamma}): $U_{\rm e}=\int N(\gamma)\gamma m_{\rm e}c^{2}d\gamma=N_{0}\gamma_{0}^{2}m_{\rm e}c^{2}\sqrt{\pi\log10/b}10^{1/4b}$ and $U_{\rm p}=m_{\rm p}c^{2}\int N(\gamma)d\gamma=N_{0}\gamma_{0}m_{\rm p}c^{2}\sqrt{\pi\log10/b}10^{1/b}$. The existence of electron-positron pairs would reduce the jet power. Note that the median value of $\gamma_{0}=1167.8$ is near the mass ratio of proton to electron, $m_{p}/m_{e}$, implying that the jet power will not reduce significantly when considering electron-positron pairs. In addition, pairs cannot largely outnumber protons, because otherwise the Compton rocket effect would stop the jet \citep[see, e.g.,][]{2010MNRAS.409L..79G}.

\begin{figure}
\begin{center}
{\includegraphics[width=1.0\linewidth]{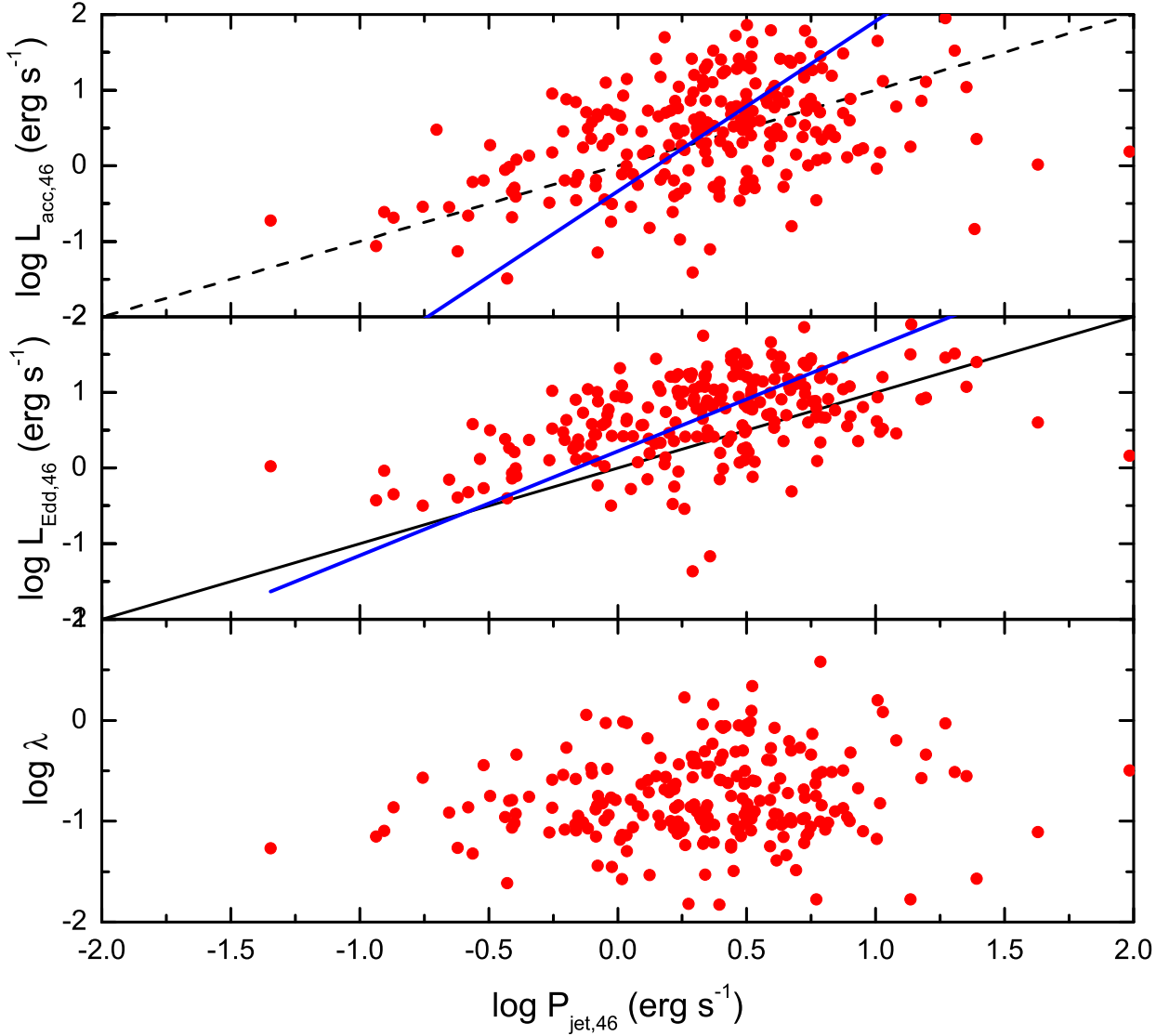}}
\end{center}
\caption{Jet power $P_{\rm jet}$ versus the accretion power $L_{\rm acc}$ (upper panel),  Eddington luminosity $L_{\rm Edd}$ (middle panel), and Eddington ratio $\lambda$ (bottom panel). The solid blue lines show robust correlations, $\log L_{\rm acc,46}=(2.25\pm0.27)\log P_{\rm jet,46}-(0.34\pm0.12)$, with a chance probability $p=2.76\times10^{-11}$ (Pearson test) for upper panel and $\log L_{\rm Edd,46}=(1.38\pm0.14)\log P_{\rm jet,46}+(0.215\pm0.064)$, with a chance probability $p=2.98\times10^{-14}$ (Pearson test) for the middle panel. The black dashed line is an equality line. There is almost no correlation between the jet power and Eddington ratio (chance probability $p=0.0354$).}
\label{fig:jet_disk}
\end{figure}

\begin{figure}
\begin{center}
{\includegraphics[width=1.0\linewidth]{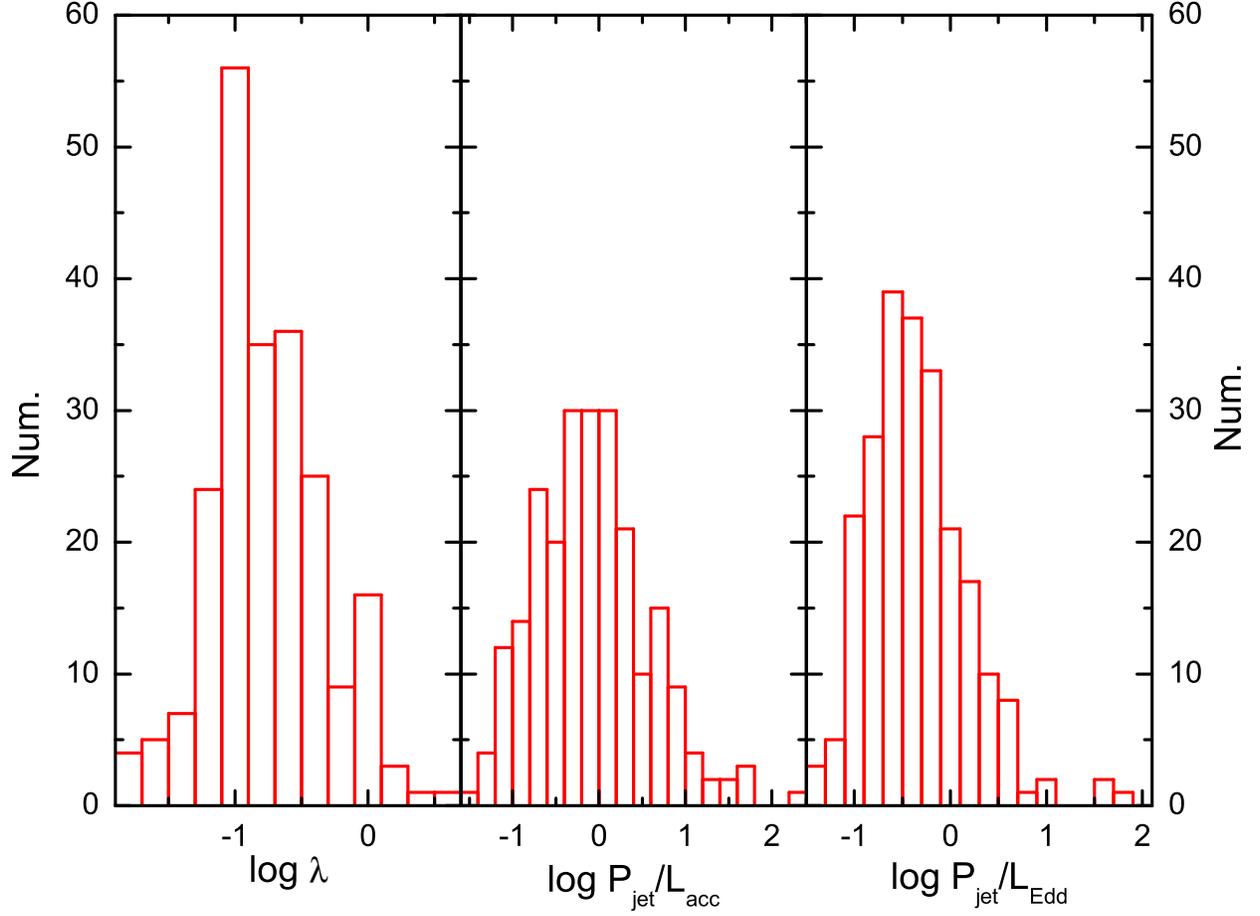}}
\end{center}
\caption{Distributions of Eddington ratio (left panel) and the jet power in units of accretion power ($P_{\rm jet}/P_{\rm acc}$, middle panel) and Eddington power ($P_{\rm jet}/L_{\rm Edd}$, right panel), of which the median values are 0.148, 0.768, and 0.382, respectively.}
\label{fig:jet_disk_distribution}
\end{figure}

In order to study the relations among the jet and accretion disk, we plot the accretion power as a function of the jet power in the Figure \ref{fig:jet_disk} (upper panel). The solid blue line shows the best-fit relation\footnote{$L_{\rm acc,46}=(L_{\rm acc}/10^{46})$ erg s$^{-1}$ and $P_{\rm jet,46}=(P_{\rm jet}/10^{46})$ erg s$^{-1}$.}: $\log L_{\rm acc,46}=(2.25\pm0.27)\log P_{\rm jet,46}-(0.34\pm0.12)$ (the Pearson test shows a chance probability $p=2.85\times10^{-11}$). The dashed black line is the equality line. This result shows that a significant number of FSRQs have jet power greater than the accretion power, supported by the distribution of the ratio of jet power to accretion power ($P_{\rm jet}/L_{\rm acc}$) as shown in the left panel of Figure \ref{fig:jet_disk_distribution}, implying that accretion power is not sufficient to launch the jets. The production of such large-power jet cannot be treated as a marginal by-product of the accretion disk flow and most likely is governed by the BZ mechanism \citep{1977MNRAS.179..433B}. The gravitational energy released from accretion cannot only be transformed into heat and radiation but also powers jets through the BP process \citep{1982MNRAS.199..883B}. In addition, the gravitational energy can also amplify the magnetic field, allowing the field to access the rotational energy of the BH and transform part of it into powerful jets through the BZ process \citep{1977MNRAS.179..433B}. As revealed by recently general relativistic magnetohydrodynamical (MHD) numerical simulations \citep[e.g.,][]{2011MNRAS.418L..79T, 2012MNRAS.423.3083M, 2014ApJ...788...71L}, the average outflowing jet/wind power can even exceed the total accretion power for the case of spin value $a=0.99$ when the BZ process dominates. The magnetic flux required to explain the production of the most powerful jets has been found to agree with the maximum magnetic flux that can be confined on BHs by the ram pressure of ``magnetically arrested disks" (MAD) \citep{2003PASJ...55L..69N}. In recent years, the MAD scenario has been thoroughly investigated and is now considered to be the likely remedy for the production of very powerful jets \citep{2011MNRAS.418L..79T, 2012MNRAS.423.3083M, 2013ApJ...764L..24S}. However, as numerical simulations suggest, the powers of the jets launched in the MAD scenario depend not only on the spin and magnetic flux but also on the disk's geometrical thickness \citep{2016MNRAS.462..636A}, with the jet power scaling approximately quadratically with all of these quantities \citep{2016Galax...4...12S}. This large power of the relativistic jet in some FSRQs suggests that the BZ process will dominate the jet launching, which confirms some previous studies \citep[see, e.g.,][]{2014Natur.510..126Z, 2014Natur.515..376G}.

The middle panel of Figure \ref{fig:jet_disk} indicates jet power as a function of Eddington luminosity, $L_{\rm Edd}=1.26\times10^{38}(M/M_{\odot})$ erg s$^{-1}$, for 229 sources with measured BH mass, which presents a significant correlation with the best linear fitting\footnote{$L_{\rm Edd,46}=(L_{\rm Edd}/10^{46})$ erg s$^{-1}$.} $\log L_{\rm Edd,46}=(1.38\pm0.14)\log P_{\rm jet,46}+(0.220\pm0.064)$ (solid blue line) and a chance probability $p=4.0\times10^{-14}$ (the Pearson test). The black dashed line is for the equality line, which shows that the jet power in units of Eddington luminosity is less than 1 for most sources. This is supported by the distribution of the ratio of jet power to Eddington luminosity, as shown in the right panel of Figure \ref{fig:jet_disk_distribution}, with a median value of 0.382.

The bottom panel of Figure \ref{fig:jet_disk} shows a relation between the Eddington ratio (defined as $\lambda=L_{\rm disk}/L_{\rm Edd}$) and jet power, suggesting almost no correlation between these two parameters \citep[the Pearson test gives $r=0.144$ and $p=0.0325$; see, e.g.,][for similar results of FSRQs]{2015ApJ...807...51Z}. This suggests that the accretion rate may not be very important for driving the jet. As discussed above, this is consistent with the idea that these jets may not be launched through the BP process, related to the accretion process, but rather through the BZ process extracting the rotational energy of the BH. Actually, BL Lacs are believed to have very weak disk continuum emission and low accretion rates compared with FSRQs \citep{2002ApJ...570L..13C, 2009ApJ...694L.107X, 2009MNRAS.396L.105G, 2014A&ARv..22...73F}. BL Lacs have comparable (very slightly lower) jet powers with FSRQs (the median values $P_{\rm jet}=$ 20.0, 6.30, and 12.0 $\times10^{45}$ erg s$^{-1}$ for FSRQs, BL Lacs, and total blazars, respectively; see discussion next subsection, the Table \ref{tab:median} and/or the bottom right panel of Figure \ref{fig:distributions}). Therefore, one may expect that the noncorrelation between Eddington ratio and jet power will extend down to the low Eddington ratio tail when considering BL Lacs.

The distribution of Eddington ratio ($L_{\rm disk}/L_{\rm Edd}$) is presented in the left panel of Figure \ref{fig:jet_disk_distribution}, with a median value of 0.148, consistent with previous results \citep[e.g.,][]{2014Natur.515..376G}. This implies that FSRQs have the standard geometrically thin, optically thick accretion disk \citep{1973A&A....24..337S}, which can launch powerful jets, contrary to some expectations \citep[e.g.,][]{1999ApJ...512..100L, 2002NewAR..46..247M}.

\subsection{Jet Magnetization and Energy Transportation}
\label{sec:power_magnetization}

The most diverse opinions about the nature of AGN relativistic jets concern their magnetic field and magnetization $\sigma=P_{\rm B}/P_{\rm kin}$, where $P_{\rm B}$ and $P_{\rm kin}$ are the magnetic and kinetic power, respectively. As discussed above, the current picture describing the powerful jet launching is that the jet can be produced from the central accretion system through the BZ mechanism by extracting BH rotation energy. This process is attributed to a key role of the dynamically important magnetic field, by means of which the BH spin energy is extracted and channeled into a Poynting/magnetic flux \citep{1977MNRAS.179..433B, 2009ApJ...699.1789T, 2011MNRAS.418L..79T}. This magnetic flux cannot be developed by dynamo mechanisms in standard radiation-dominated accretion disks \citep{1997MNRAS.292..887G}. However, such a flux is expected to be accumulated in the inner regions of the accretion flow and/or on the BH by the advection of magnetic fields from external regions \citep[see][and references therein]{2013ApJ...764L..24S, 2013ApJ...765..149C}. In fact, the magnetic flux close to the BH horizon is so large that accretion likely occurs through a MAD flow, as discussed above \citep{2003PASJ...55L..69N, 2011MNRAS.418L..79T, 2012MNRAS.423.3083M}. Starting as Poynting flux-dominated outflows, the jets are smoothly accelerated as magnetic power is being progressively converted to kinetic power and their magnetization drops, until a substantial equipartition between the magnetic and the kinetic power is established \citep[$\sigma\approx1$; see, e.g.,][and references therein]{2007MNRAS.380...51K, 2009ApJ...699.1789T, 2015ASSL..414..177V}. Therefore, at the end of this acceleration phase, the jet should still carry a substantial fraction ($\approx$half) of its power in the form of a Poynting flux. The jet acceleration becomes inefficient when $\sigma\lesssim1$, from which point on, in the ideal MHD picture, $\sigma$ will decrease logarithmically \citep{2010MNRAS.402..353L}.

However, even though blazars can have jets originating from MADs and be powered by the BZ mechanism, at the same time, they can have $\sigma\ll1$ at the radio core and/or the blazar zone \citep[i.e., the region where most of the radiation is produced, as indicated by blazar models with jet opening angles $1/\Gamma$;][]{2014ApJ...796L...5N, 2015MNRAS.449..431J, 2015MNRAS.451..927Z}. \citet{2015MNRAS.451..927Z} suggested that the magnetic-to-kinetic energy flux conversion is assumed to result from the differential collimation of poloidal magnetic surfaces, which is the only currently known conversion mechanism in steady-state, axisymmetric, and nondissipative jets\footnote{If there is a dissipative mechanism, the magnetic energy could be converted to thermal and kinetic energy.} \citep[the conversion process can initially proceed quite efficiently; see also][]{2009ApJ...699.1789T, 2010MNRAS.402..353L}. Hence, it is likely that other mechanisms are involved in the conversion process working at $\sigma\lesssim1$, such as MHD instabilities \citep[see][and references therein]{2011MmSAI..82...95K}, randomization of magnetic fields \citep{2000ApJ...535..104H}, reconnection of magnetic fields \citep{2002A&A...391.1141D, 2010ApJ...725L.234L} and/or impulsive modulation of jet production \citep{2010ApJ...722..197L, 2011MNRAS.411.1323G}. Little is known about the feasibility and efficiency of these processes in the context of AGNs. The MHD instabilities can develop when $\sigma$ drops to unity or even earlier if stimulated by high-amplitude fluctuations of the jet power and direction, which are predicted in the MAD model \citep[e.g.,][]{2012MNRAS.423.3083M}. In this case, the magnetization can reach $\sigma\ll1$ even prior to the radio core and/or the blazar zone.

Studies of the $\sigma$ parameter are important, not only for better understanding of the dynamical structure and evolution of relativistic jets but also because its value determines the dominant particle acceleration mechanism and its efficiency. Dissipation of part of the kinetic (through shocks) and/or magnetic (through reconnection) power leads to the acceleration of particles up to ultrarelativistic energies, producing the nonthermal emission we observe from blazars. Both versions (the first/second order) of Fermi acceleration were, and still are, commonly invoked as responsible for the ultrarelativistic electrons in AGN jets \citep[see, e.g.,][]{1992ApJ...390..454H, 2011ApJ...726...75S, 2014ApJ...780...87M}. Based on so-called particle-in-cell simulations, it is shown that electrons can be efficiently heated in very low magnetized jets \citep[$\sigma\ll1$;][]{2016ARA&A..54..725M}. The reconnection of magnetic energy is also a promising mechanism for particle acceleration in AGN jets \citep[e.g.,][]{2009MNRAS.395L..29G, 2011SSRv..160...45U, 2012ApJ...754L..33C}. A prediction of this scenario, resulting from detailed particle-in-cell simulations, is the substantial equipartition between the magnetic field and the accelerated electrons downstream of the reconnection site, where particles cool and emit the radiation we observe. For example, relativistic reconnection 2D models predict a lower limit of the order of $\sigma\sim0.3$ in the dissipation region \citep{2015MNRAS.450..183S}. Needless to say, the magnetic reconnection scenario requires that jets carry a sizable fraction of their power in the magnetic form up to the emission regions.

\begin{figure}
\begin{center}
{\includegraphics[width=1.0\linewidth]{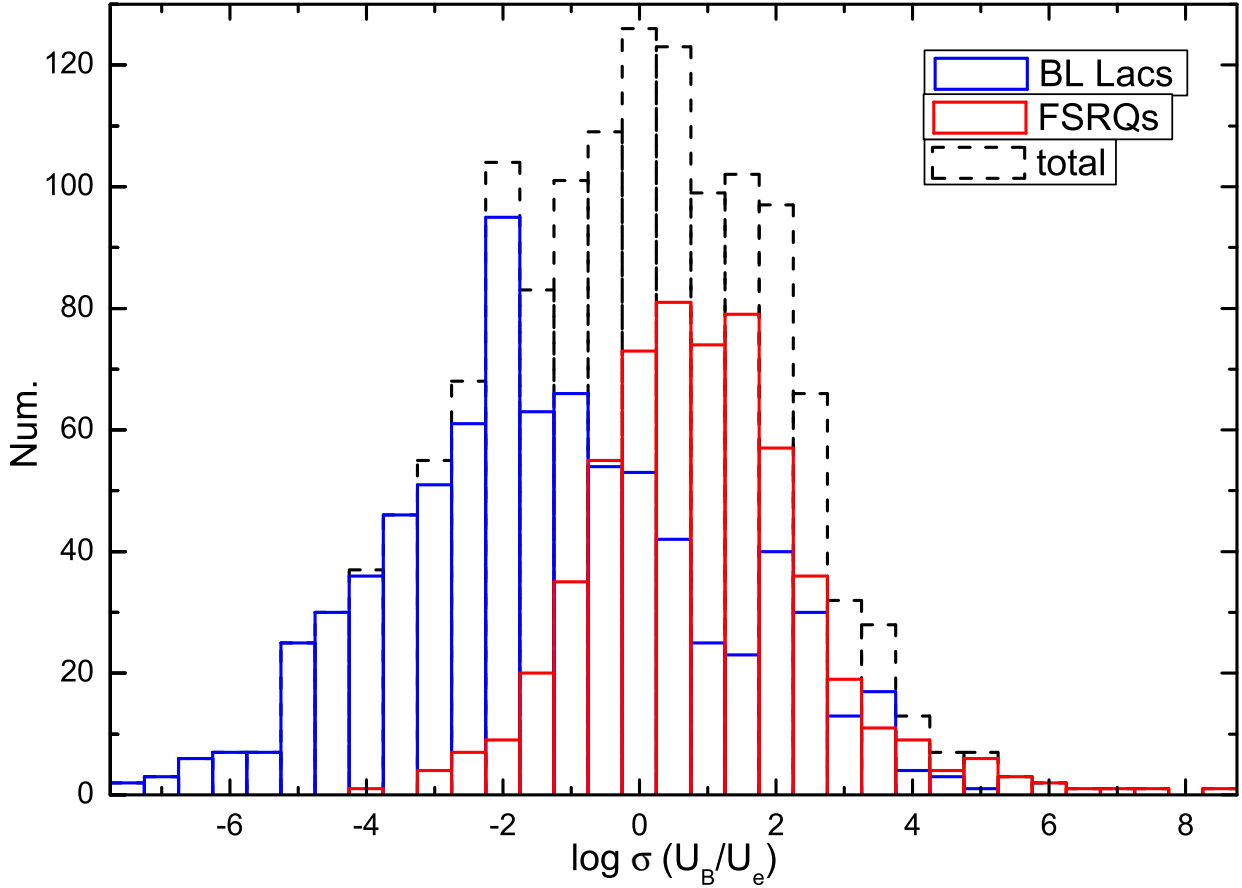}}
\end{center}
\caption{Distribution of the magnetization parameters $\sigma=U_{\rm B}/U_{\rm e}$ of a total of 1392 blazars with median values of FSRQs, BL Lacs, and total blazars of 6.42, 0.0285, and 0.640, respectively. The KS test yields the significance level probability for the null hypothesis that FSRQs and BL Lacs are drawn from the same distribution $P=4.67\times10^{-76}$ and the statistic $D_{\rm KS}=0.503$ (the maximum separation of the two cumulative fractions).}
\label{fig:sigma}
\end{figure}

Based on the jet parameters estimated in this paper, the magnetization properties can be studied. Here we define a magnetization parameter as the ratio of energy densities between the magnetic field and relativistic electrons\footnote{Note that the definition of $\sigma=U_{\rm B}/U_{\rm e}$ here is different from $P_{\rm B}/P_{\rm kin}=U_{\rm B}/(U_{\rm e}+U_{\rm p})$. For typical jet parameters (median values), $U_{\rm p}/U_{\rm e}=10^{3/4b}(m_{p}/m_{e})/\gamma_{0}\approx28.0$, 3.51, and 11.6 for FSRQs, BL Lacs, and total blazars.} $\sigma=U_{\rm B}/U_{\rm e}$ (also called the equipartition parameter). The distribution of $\sigma$ is shown in the right panel of Figure \ref{fig:sigma}. It can be seen that FSRQs have a narrower distribution than that of BL Lacs, with a median values $\sigma=$6.42 and 0.0285 (and 0.640 for total blazars), respectively (see also Table \ref{tab:median}). The Kolmogorov-Smirnov (KS) test yields the significance level probability for the null hypothesis that FSRQs and BL Lacs are drawn from the same distribution $p=4.67\times10^{-76}$ and the statistic $D_{\rm KS}=0.503$ (the maximum separation of the two cumulative fractions). Similar results have been presented on single sources \citep{2011ApJ...736..131A, 2011ApJ...738...25A, 2015MNRAS.451..739A} and through modeling the SEDs of the large sample blazars \citep{2013ApJ...774L...5Z, 2016MNRAS.456.2374T}, which demonstrate that the magnetization parameter of the large majority of BL Lacs is commonly at $\sim10^{-2}$. \citet{2017arXiv171106282C} studied six hard-TeV BL Lacs and found that the equipartition parameters range from $10^{-3}$ to $10^{-5}$. \citet{2012ApJ...748...81K} presented a deep Chandra/ACIS observations and Hubble Space Telescope (HST) Advanced Camera for Surveys observations of two FSRQs, PKS B0106+013 and 3C 345, and found the shocked jet regions upstream of the radio hot spots, the kiloparsec-scale jet wiggles, and a ``nose cone"-like jet structure in PKS B0106+013, as well as the V-shaped radio structure in 3C 345, which suggest that the jet still has a large $\sigma$ at these large scale. Modelling of FSRQ SEDs is also in agreement with this result \citep[e.g.,][]{2015MNRAS.448.1060G}.

The smaller $\sigma$ in BL Lacs relative to FSRQs implies that BL Lacs jets would suffer deceleration more easily than those of FSRQs \citep{2008ApJ...678...64P, 2015ApJ...798..134H}, although both present similar violent $\gamma$-ray emission from the central region. This is consistent with the idea that some BL Lacs show almost no superluminal motion in the Very Long Baseline Array (VLBA) scale \citep[e.g., Mrk 501 and Mrk 421;][]{2004ApJ...600..127G, 2006ApJ...646..801G}, while the multiwavelength SED and variability imply a highly Doppler-booting emission, which indicates that the jet has already been decelerated at that scale \citep[e.g.,][]{2005A&A...432..401G, 2004ApJ...613..752G}. For some FSRQs \citep[e.g., 3C 279 and 3C 345;][]{2003ApJ...588..716P, 1999ApJ...521..509L}, the Monitoring of Jets in AGN with VLBA Experiments (MOJAVE) project shows that jets will still be accelerated at the VLBA scale \citep[see,][]{2015ApJ...798..134H}. The low magnetization in BL Lacs is essential in order to avoid efficient cooling of relativistic electrons, which would lead to much larger values of $\gamma_{0}$ of BL Lacs relative to FSRQs (see Table \ref{tab:median} and/or Figure \ref{fig:distributions}). This also seems to  imply that the magnetic reconnection cannot be strong in these emission regions. The particle acceleration in situ may originate from the energy dissipated in shocks and/or in boundary shear layers. The spine/layer jet structure has been revealed through many observational techniques and can contribute significant X-ray emission due to the enhancement of the seed photon energy density by relative opposite motions between the spine and layer \citep[see][for recent review]{2017ApJ...842..129C}.

So far, there is no observational evidence supporting the theoretical prediction that a high-magnetization jet will more easily to transport energy to a large scale than a low-magnetization jet. As found above, FSRQs have relatively larger values of $\sigma$ than BL Lacs, which simply predicts that the jet may transport energy more efficiently to a large scale in FSRQs than BL Lacs. Therefore, we may expect a larger extended jet kinetic power (relative to the core SED jet power) in FSRQs than BL Lacs. To answer this question, we should estimate the (extended) jet kinetic power of these sources.

The jet power is basically important for understanding jet formation, the jet-disk relation \citep{1977MNRAS.179..433B, 1982MNRAS.199..883B, 2009MNRAS.396..984G, 2010LNP...794..233S, 2012MNRAS.419L..69N, 2012ApJ...753...24L, 2013ApJ...770...31W, 2014ApJ...783...51C}, and the AGN feedback on structure formation \citep{2007MNRAS.379..260M, 2007MNRAS.379..894B}. It is generally believed that the X-ray cavities are direct evidence of AGN feedback and provide a direct measurement of the mechanical energy released by the AGN jets (jet kinetic power) through the work done on the hot, gaseous halos surrounding them \citep[e.g.,][]{2004ApJ...607..800B, 2006MNRAS.372...21A, 2010ApJ...720.1066C}. It is also found that the cavity kinetic power is correlated with the radio extended power \citep{1991Natur.349..138R, 2004ApJ...607..800B, 2010ApJ...720.1066C}, and the scaling relationship is roughly consistent with the theoretical relation \citep[e.g.,][]{1999MNRAS.309.1017W, 2010ApJ...720.1066C, 2011ApJ...740...98M}. Therefore, in the systems where the X-ray cavities are lacking or not observed, the radio extended luminosity is used to estimate the jet kinetic power.
The relations presented in \citet{1999MNRAS.309.1017W} are commonly used to do this estimation \citep{2010ApJ...720.1066C}. Based on this method, \citet{2011ApJ...740...98M} presented the largest blazar sample with measured (extended) jet kinetic power, until now.

\begin{figure}
\begin{center}
{\includegraphics[width=1.0\linewidth]{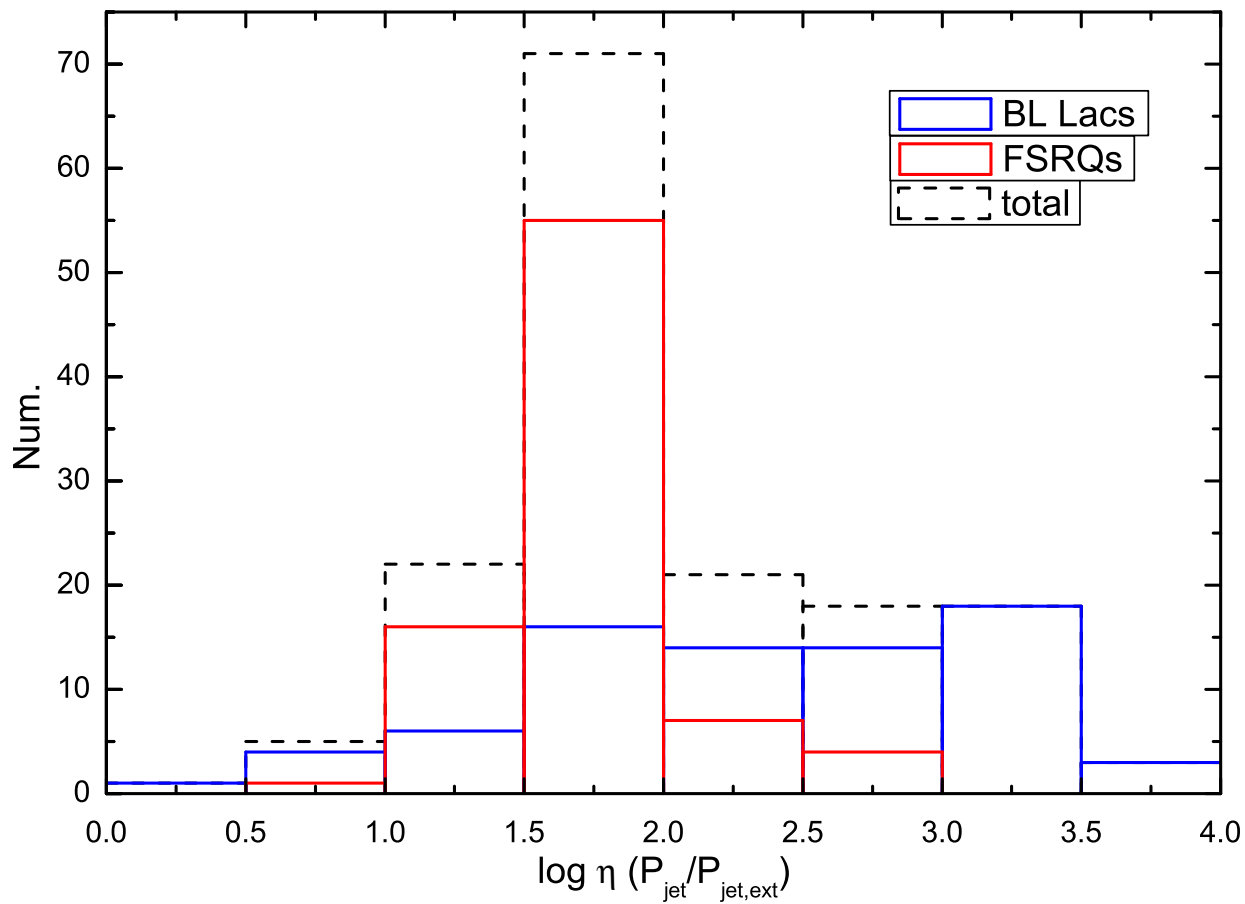}}
\end{center}
\caption{Distribution of $\eta=P_{\rm jet}/P_{\rm jet,ext}$ for blazars having measured extended radio emission (159 blazars, including 83 FSRQs and 76 BL Lacs), with median values of 57.2, 230, and 73.9 for FSRQs, BL Lacs, and total blazars, respectively. The KS test gives the significance level probability that FSRQs and BL Lacs are drawn from the same distribution $P=2.58\times10^{-10}$ and the statistic $D_{\rm KS}=0.524$.}
\label{fig:jetpower_ext_distribution}
\end{figure}

\begin{figure}
\begin{center}
{\includegraphics[width=1.0\linewidth]{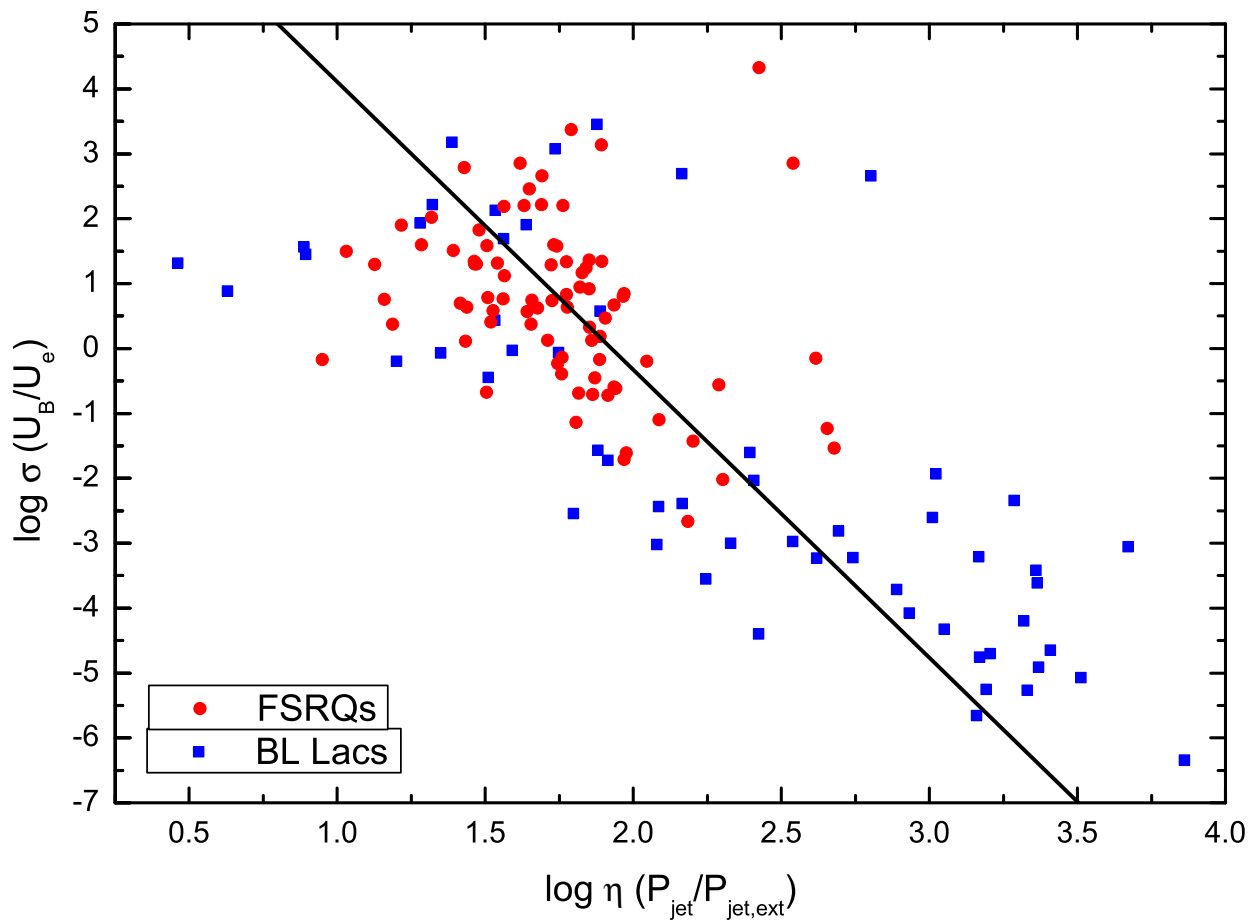}}
\end{center}
\caption{Jet magnetization parameters ($\sigma=U_{\rm B}/U_{\rm e}$) versus the ratio of (SED) jet power to (extended) kinetic jet power ($\eta=P_{\rm jet}/P_{\rm jet,ext}$). The solid black line presents a best linear fit for the total sources, $\log\sigma=-(4.44\pm0.33)\log\eta+(8.54\pm0.69)$, with a chance probability $p=1.23\times10^{-26}$ (Pearson test).}
\label{fig:eta_sigma}
\end{figure}

In this paper, we collect the (extended) jet kinetic power, $P_{\rm jet,ext}$, from \citet{2011ApJ...740...98M}, resulting in a total of 159 blazars, including 76 BL Lacs and 83 FSRQs, whose data are presented in Table \ref{tab:data}. To study the jet energy transportation, we define $\eta$ as a ratio of the jet SED power (from the central nuclei, $P_{\rm jet}$) to the (extended) jet kinetic power ($P_{\rm jet,ext}$), $\eta=P_{\rm jet}/P_{\rm jet,ext}$. The distribution of $\eta$ is shown in Figure \ref{fig:jetpower_ext_distribution} with median values $\eta=$57.2, 230, and 73.9 for FSRQs, BL Lacs, and total blazars, respectively, which implies that the jet SED power is significantly larger than the (extended) jet kinetic power. Two reasons may account for this: (1) the (SED) jet power may represent jet power measured in AGN active state, while the (extended) jet kinetic power is the historically average power, and 2) most of the energy is dissipated at the central region before being transported to a large scale. We note that FSRQs have an average smaller $\eta$ than BL Lacs (by about one order of magnitude). The KS test gives the significance level probability that FSRQs and BL Lacs are drawn from the same distribution $P=2.58\times10^{-10}$ and the statistic $D_{\rm KS}=0.524$. This suggests that relatively more energy is transported to a large scale in FSRQs than in BL Lacs. As discussed above, the central jets of FSRQs also have relatively higher magnetization ($\sigma$) than those of BL Lacs, suggesting that a higher-magnetization jet can more easily transport energy to a large scale. As discussed above, some BL Lacs have extremely estimated Doppler factors. We instead use the median value of $\delta$ instead to determine the other jet parameters for these sources. This uncertainty may affect our results. Therefore, we exclude all BL Lacs with $\delta>100$ or $\delta<1$ and find that the remaining 56 BL Lacs have median values of $\eta=230$ and $\sigma=0.00389$, which is also consistent with the above result that the higher-magnetization jet can more easily transport energy to a large scale. For further research, we plot $\eta$ versus $\sigma$ in Figure \ref{fig:eta_sigma} for sources with $1<\delta<100$, which presents a significant anticorrelation. The solid black line shows the best linear fit, $\log\sigma=(8.54\pm0.69)-(4.44\pm0.33)\log\eta$, with a chance probability $p=1.23\times10^{-26}$ (Pearson test). This result confirms the above result, suggesting that a higher-magnetization jet can more easily transport energy to a larger scale, which, for the first time, offers supporting evidence for the jet energy transportation theory.

Our sample include 393 unknown-redshift sources (157 BL Lacs and 236 BCUs), which is about 28\% of the sample. We use the median values of known redshifts of blazars for these sources. In order to explore the possible effect of this replacement on our results, we recalculate the median values of all jet parameters of another two cases: the case 1 uses known-redshift sources including BCUs, and the case 2 uses known-redshift sources excluding BCUs. All of these values are presented in the Table \ref{tab:median}, where the label ({\it z}) is for the case 1 and the label ({\it cz}) is for the case 2. It can be seen that both cases are consistent with our above results.

\section{Conclusion}
\label{sec:conclusion}

In this paper, based on broadband SEDs, we estimate the jet physical parameters of 1392 $\gamma$-ray-loud AGNs \citep{2015ApJ...810...14A, 2016ApJS..226...20F}. These values are presented in the Table \ref{tab:data}, and the median values are shown in the Table \ref{tab:median}, which are roughly consistent with previous studies. Out of these sources, the accretion disk luminosities of 232 sources and the (extended) kinetic jet powers of 159 sources are compiled from archived papers. The full version of the data is available online (publicly). The main results are summarized here.

1. We show that $\gamma$-ray FSRQs and BL Lacs are well separated by $\Gamma=-0.127\log L_{\gamma}+8.18$, in the $\gamma$-ray luminosity versus photon index plane with a success rate of 88.6\%. This criterion is employed to divide 311 BCUs into FSRQs or BL Lacs.

2. The peak electron energy $\gamma_{0}$ forms two distinct distributions between FSRQs and BL Lacs (also for $\nu_{\rm s}$/$\nu_{\rm C}$ distributions), implying that the electrons in FSRQ jets suffer strong cooling, while they suffer less cooling in BL Lac jets.

3. A Significant number of FSRQs present a jet bolometric luminosity larger than disk luminosity by $\sim2$ orders of magnitude, which implies a (SED) jet power (assuming typical values of the Doppler factor $\delta\sim10$ and jet radiative efficiency $\sim10\%$) larger than the disk luminosity by $\sim$10 times. This suggests that the jet-launching processes and the way of transporting energy from the vicinity of the BH to infinity must be very efficient.

4. Most FSRQs present a (SED) jet power larger than the accretion power, which suggests that the relativistic jet-launching mechanism is dominated by the BZ process, at least in these FSRQs.

5. The magnetization of jets in BL Lacs is significantly lower than that in FSRQs, which is consistent with the idea that BL Lac jets may be more easily decelerated than FSRQ jets.

6. The ratio of the (extended) kinetic jet power to the (SED) jet power in FSRQs is significantly larger than that in BL Lacs. There is a significant anticorrelation between the jet magnetization parameter and the ratio of the SED jet power to the (extended) kinetic jet power. These results, for the first time, provide supporting evidence for the jet energy transportation theory: a high-magnetization jet can more easily transport energy to a large scale than a low-magnetization jet.

It should be noted that the nonsimultaneity of the SED does not always have an optimal impact on the derived parameters (and therefore large uncertainties) for single sources, but this is compensated for by the large number of sources ensuring that, from the statistical point of view, they should be only statistically meaningful.

\acknowledgments

We thank the anonymous referee for insightful comments
and constructive suggestions. We are grateful for help from Xinwu Cao, Shiyin Shen, and Zhaoming Gan. This work is supported by the CAS grant (QYZDJ-SSW-SYS023).

\begin{table}
\tabletypesize{\tiny}
\setlength{\tabcolsep}{4.0pt}
\caption{AGN Jet physical parameters}
\begin{tabular}{@{}lcccccccccccccc@{}}
\hline
\hline
Name & $z$ & Cl. & $\log\gamma_{0}$ & $b$ &  $\log R$  &  $\delta$ & $\log B$ &  $\log P_{\rm jet}$ & $\log\sigma$ & $\log\eta$ & $\log M$ & $\log L_{\rm disk}$ & Model\\
(1)  &  (2)  &  (3)  &  (4) &  (5)& (6) & (7) &  (8)  &  (9)  &  (10) & (11) & (12) & (13) & (14)\\
\hline
J0001.2-0748    &     -     &   CB   &    4.25   &    0.60   &   17.2   &   56.1   &   -2.46   &   46.6   &   -3.36   &     -     &     -     &     -     &   ST   \\
......\\
\hline
\end{tabular}
\\
{\bf Note.}
The source names labeled with ``$\ast$" are blazars having extreme Doppler factors ($\delta<1$ or $\delta>100$), and therefore we instead use the median values ($\delta_{\rm m}\simeq14.3$ for BL Lacs and $\delta_{\rm m}\simeq10.7$ for FSRQs) instead to calculate other jet parameters (see text for details). Column (1) gives the 3FGL name. Column (2) gives the redshift. Column (3) is the class of the source: ``UF" for BCUs classified as FSRQs, ``UB" for BCUs classified as BL Lacs, and ``CF" and ``CB" for confirmed FSRQs and BL Lacs, respectively. Column (4) is the electron peak energy. Column (5) is the curvature $b$ of the electron energy distribution. Column (6) is the radius of the emission sphere in units of cm. Column (7) is the Doppler factor. Column (8) is the strength of the magnetic field in units of Gs. Column (9) is the jet (SED) power in units of erg s$^{-1}$. Column (10) is the magnetization parameter $\sigma=U_{\rm B}/U_{\rm e}$. Column (11) is the ratio of the jet SED power to (extended) jet kinetic power, $\eta=P_{\rm jet}/P_{\rm jet,ext}$. Column (12) is the available BH mass in units of $M_{\odot}$. Column (13) is the available disk luminosity in units of erg s$^{-1}$. Column (14) is the model used in the calculations: ``ST" for SSC/Thomson, ``SK" for SSC/KN, ``ET" for EC/Thomson, and ``EK" for EC/KN.\\
(This table is available in its entirety in ASCII text form.)
\label{tab:data}
\end{table}

\begin{table}
\tabletypesize{\tiny}
\setlength{\tabcolsep}{4.0pt}
\caption{Median values of Some jet physical parameters}
\begin{tabular}{@{}lcccccccccccc@{}}
\hline
\hline
     & $\gamma_{0}$ & $b$ &  $R$  &  $\delta$ & $B$ &  $P_{\rm jet}$ & $\eta$ & $\sigma$ & $P_{\rm jet}/L_{\rm acc}$ & $P_{\rm jet}/L_{\rm Edd}$ & $L_{\rm disk}/L_{\rm Edd}$\\
     & (1)  &  (2)  &  (3)  &  (4) &  (5)& (6) & (7) &  (8)  &  (9)  &  (10) & (11) \\
\hline
FSRQ({\it t})   & 1167.8 & 0.6  & 2.78 & 10.7 & 1.56  & 20.0 & 57.2  & 6.42   & 0.768 & 0.382 & 0.148\\
BL Lac({\it t}) & 12077  & 0.55 & 3.70 & 14.3 & 0.119 & 6.3  & 230   & 0.0285 & - & - & -\\
Total({\it t})  & 3646.1 & 0.55 & 3.70 & 14.3 & 0.446 & 12.0 & 73.9  & 0.640  & - & - & -\\
\hline

FSRQ({\it z})   & 1075.9 & 0.65 & 2.93 & 11.3 & 1.54   & 22.2 & 57.2  & 6.42   & 0.768 & 0.382 & 0.148\\
BL Lac({\it z}) & 13888  & 0.55 & 3.70 & 14.3 & 0.0748 & 5.08 & 230   & 0.0121 & - & - & -\\
Total({\it z})  & 2713.3 & 0.60 & 3.70 & 14.3 & 0.528  & 14.8 & 73.9  & 0.743  & - & - & -\\
\hline

FSRQ({\it cz})   & 1050.8 & 0.65 & 2.96 & 11.4 & 1.54   & 22.2 & 57.2 & 6.43   & 0.768 & 0.382 & 0.148\\
BL Lac({\it cz}) & 14036  & 0.55 & 3.70 & 14.3 & 0.0768 & 5.11 & 213  & 0.0131 & - & - & -\\
Total({\it cz})  & 2614.6 & 0.60 & 3.70 & 14.3 & 0.554  & 15.3 & 73.4 & 0.862  & - & - & -\\

\hline
\end{tabular}
{\bf Note.} Case ({\it t}) is for all sources in our sample. Case ({\it z}) is for sources having known redshfits and including BCUs, while case ({\it cz}) is for sources having known redshfits and excluding BCUs. Column (1) is the electron peak energy. Column (2) is the curvature of the electron energy distribution. Column (3) is the radius of the emission sphere in units of $10^{16}$ cm. Column (4) is the Doppler factor. Column (5) is the strength of the magnetic field in units of Gs. Column (6) is the jet (SED) power in units of $10^{45}$ erg s$^{-1}$. Column (7) is the ratio of the jet SED power to the (extended) jet kinetic power, $\eta=P_{\rm jet}/P_{\rm jet, ext}$. Column (8) is the magnetization parameter $\sigma=U_{\rm B}/U_{\rm e}$. Column (9) is the jet power in units of accretion power. Column (10) is the jet power in unit of Eddington power. Column (11) is the disk luminosity in units of Eddington power.
\label{tab:median}
\end{table}

\end{document}